\documentclass[preprint,nofootinbib,superscriptaddress,amssymb]{revtex4}
\pdfoutput=1
\usepackage{amsmath} 
\usepackage{verbatim} 
\usepackage{graphicx} 
\usepackage{subfig} 
\usepackage{color}
\usepackage{ulem}

\begin{document}

\title{How Cold is Cold Dark Matter?}

\author{Cristian Armendariz-Picon}
\affiliation{Department of Physics, Syracuse University, Syracuse, NY 13244-1130, USA}

\author{Jayanth T. Neelakanta}
\affiliation{Department of Physics, Syracuse University, Syracuse, NY 13244-1130, USA}

\begin{abstract}
If  cold dark matter consists of particles,  these must be non-interacting and non-relativistic by definition. In most cold dark matter models however, dark matter particles inherit a non-vanishing velocity dispersion from interactions in the early universe, a velocity that redshifts with cosmic expansion but certainly remains non-zero.  In this article, we place model-independent constraints on the  dark matter temperature to mass ratio, whose square root determines the dark matter velocity dispersion. We only assume that dark matter particles decoupled kinetically while non-relativistic, when galactic scales had not entered the horizon yet, and that their momentum distribution has been Maxwellian since that time. Under these assumptions, using cosmic microwave background and matter power spectrum observations, we place upper limits on the temperature to mass ratio of cold dark matter today (away from collapsed structures). These limits  imply that the present cold dark matter velocity dispersion  has to be smaller than 54~m/s.  Cold dark matter has to be quite cold, indeed.

\end{abstract}

\maketitle

\section{Introduction}

A wide array  of observations, ranging from the distribution of matter on cosmological distances, to the rotation curves of galaxies on kiloparsec scales, suggest that the universe contains a form of matter that does not interact with electromagnetic radiation and whose pressure is negligible.   At present, the nature of this dark matter is  unknown, but, among other hints,  the former phenomenological properties  strongly suggest that dark matter is made of non-relativistic particles that couple very weakly to the standard model. 

There is certainly no shortage of  particle dark matter models accommodating these properties, such as axions, moduli, gravitinos, Kaluza-Klein excitations, sterile neutrinos or WIMPs, just to name a few \cite{Feng:2010gw}. In each different model, dark matter experiences  a different  cosmic evolution, resulting in a distribution of dark matter momenta  that is often thermal at late times, although with temperatures that span many orders of magnitude in the different dark matter scenarios.  Given the great variety of dark matter models and associated thermal histories, it is thus natural to ask whether we can place phenomenological limits on the dark matter temperature today, or whether in fact there is evidence that dark matter has a non-zero temperature. Indeed, several authors have suggested that warm dark matter  (see below) may help alleviate the apparent tension between the predictions of the Cold-Dark-Matter (CDM) scenario and  the actual amount of clustering on sub galactic scales \cite{Colombi:1995ze, SommerLarsen:1999jx, Colin:2000dn, Bode:2000gq,Haiman:2001dg, deVega:2011si}, although recent studies suggest that  warm dark matter  is disfavored by observations \cite{Viel:2013fqw}. Similarly, the same small-structure problems may also be avoided if  cold dark matter  kinetically decouples rather late in cosmic history, as described in  \cite{Aarssen:2012fx} and references therein. Phenomenological bottom-up  limits on CDM like the one we discuss would not only constrain  many of the different cold dark matter models, but also offer us a generic model-independent way to further characterize the properties of dark matter. 

In this article we explore model-independent limits on the dark matter temperature to mass ratio extrapolated to the present time, $T_0/m$. This ratio determines the velocity dispersion of dark matter particles, which is a parameter that directly controls the growth of  structure.  If  dark mater  decouples while non-relativistic, dark matter particles travel at a root-mean-square velocity $v_\mathrm{rms}=\sqrt{3T/m}$, where $m$ is the dark matter mass. As a result,  anisotropies on scales much smaller than the associated free-streaming length are strongly suppressed, a phenomenon usually known as Landau damping. The absence of such suppression on observable scales thus places limits on the dark matter temperature to mass ratio.   Whereas it turns out to be convenient to frame our limits in terms of the dark matter temperature, they can be equally interpreted as limits on the root-mean-square dark matter velocity. In this context, we should also point out that our limits apply even if dark matter  does not consist of  elementary point particles, but, instead, is made of objects of even macroscopic size, provided that their velocity dispersion is Maxwellian.

In order to reliably calculate the impact of a non-zero dark matter temperature on the formation of structure, we restrict ourselves to linear perturbation theory, and thus focus on cosmological probes applicable in this regime: the CMB and the matter power  spectrum on the appropriate scales.   The same suppression of structure implied by a non-zero temperature also impacts the smallest scales that become non-linear, and, thus, the mass of the smallest proto-halos. Several authors have studied how the latter depend on the mass parameters of specific dark matter models, such as WIMPs in supersymmetric or extra-dimensional models, but in these cases the corresponding scales are too small  to be probed observationally \cite{Green:2005fa, Moore:2005uu, Profumo:2006bv}. Interactions between dark matter particles and the thermal bath in the early universe may also imprint features on the matter power spectrum on sub-horizon scales at the time of kinetic decoupling \cite{Loeb:2005pm, Bertschinger:2006nq}.  These interactions may lead to a suppression of small scale structure  that is stronger than  that due to free streaming on  those scales, although they do not have any impact on modes outside the horizon at that time.  Because the data used in our analysis  only probe modes that entered the horizon at $z<z_\mathrm{max}\approx 5\cdot 10^5 $, these features are absent in the modes of interest if kinetic decoupling occurred at $z_\mathrm{dec}>z_\mathrm{max}$, an assumption typically satisfied in most dark matter models. In fact,  a recent analysis of cosmological data does not find any evidence for  features due to dark matter interactions \cite{Cyr-Racine:2013fsa}, further suggesting that cosmological scales must have been outside the horizon at the time of kinetic decoupling. In that case, and in the context of our analysis, we can simply assume that cold dark matter is effectively collisionless. Signatures of  dark matter interactions are then buried in small scales, beyond the reach of our cosmological data.

The ratio $T_0/m$, and any quantity derived from it,  should be carefully interpreted. By the  former we mean the temperature to mass ratio dark matter would have today in the absence of structure formation. In the real universe, however, dark matter inhomogeneities grow, become non-linear and collapse,  resulting in virialized dark matter haloes whose temperature  is determined just by the properties of the halo. In pure cold dark matter models the Press-Schechter mass function predicts that all of the dark matter ends up in such  halos \cite{Press:1973iz, Bond:1990iw}. If the dark matter temperature is non-zero there is a cut-off in the matter power spectrum at small scales, implying that only a fraction of dark matter collapses, but we expect this fraction to be significant for small enough temperatures.  Hence, our ratio $T_0/m$ is not the temperature of a typical dark matter particle in today's universe, but just an extrapolation of what that ratio would be had dark matter not collapsed. 

The work in the literature  closest  to the limits we discuss here  has mainly   focused on constraints  on the mass of  warm dark matter particles  \cite{Abazajian:2005xn,Viel:2005qj,Boyarsky:2008xj,deSouza:2013wsa,Viel:2013fqw}. In these models,  dark matter is  typically assumed to be hot, in the sense that it decouples kinetically while being relativistic, whereas in our work we assume that dark matter is cold, and thus decouples while non-relativistic. Because the parameter that actually affects the formation of dark matter structure is the velocity dispersion, in order to relate the latter to  the dark matter mass, previous analyses  have typically needed to introduce additional assumptions tied to the particular dark matter model being considered, such as the chemical potential of dark matter, its thermal history,   or its number of degrees of freedom.  As a consequence,  mass limits only apply strictly in the context of the models in which they were derived, and cannot be readily extended to other scenarios. The present  work focuses instead on the dimensionless ratio of temperature to dark matter mass directly, and basically relies on  just  two assumptions: that the distribution of dark matter momenta is Maxwellian after the time the smallest relevant scales entered the horizon, and that dark matter has been collisionless at least since that time. 

\section{Formalism}
\label{sec:Formalism}
We assume that dark matter consists of collisionless particles, which for simplicity and without  loss of generality we take to be spinless. As we pointed out in the introduction, these particles do not have to be point-like: As far as our analysis is concerned, the only restriction is that their size be much smaller than any other length scale in the problem.
 
Under these conditions,   dark matter  is then characterized by its distribution function $f$, which counts the number density of particles at coordinate time $\tau$, comoving coordinate $\vec{x}$ and covariant momentum $\vec{p}$. Since we assume these particles to be collisionless,  their distribution function $f(\tau, x^i, p_j)$ obeys the  collisionless Boltzmann equation
\begin{equation}\label{eq:Boltzmann NC}
	\frac{p^0}{m}\left[\frac{\partial f}{\partial \tau}
	+\frac{\partial f}{\partial x^i}\frac{dx^i}{d\tau}
	+\frac{\partial f}{\partial p_i}\frac{dp_i}{d\tau}\right]=0,
\end{equation}
in which the zero on the right hand side accounts for the absence of non-gravitational interactions. 

Because we assume that dark matter is non-interacting, its only observable effects involve gravitation.  The energy momentum tensor of a distribution of dark matter particles characterized by $f$ is
\begin{equation}\label{eq:EMT}
	T_{\mu\nu}= \frac{1}{\sqrt{-g}}\int d^3 p \, \frac{ p_\mu p_\nu}{p^0} f,
\end{equation}
where $g$ is the determinant of the space-time metric.  
 
\subsection{Background Distribution}
In a spatially flat  Friedman-Robertson-Walker universe with space-time metric  
\begin{equation}\label{eq:FRW}
	ds^2=a^2(\tau)\left[-d\tau^2+d\vec{x}^2\right],
\end{equation}
the Boltzmann equation (\ref{eq:Boltzmann NC}) reads
\begin{equation}\label{eq:Boltzmann background}
	\frac{\partial f}{\partial \tau}=0.
\end{equation}
Hence, any homogeneous and isotropic distribution $f=f(p)$,  where
\begin{equation}\label{eq:p}
	p\equiv a\sqrt{g^{ij} p_i p_j},
\end{equation}
is a solution of the Boltzmann equation (\ref{eq:Boltzmann background}). Because we want to describe cold dark matter, we assume that the distribution function describes a gas of non-relativistic particles, and therefore choose it to be the Maxwell-Boltzmann distribution
\begin{equation}\label{eq:MB}
	f(p)=\frac{1}{(2\pi)^3}\exp\left(-\frac{m-\mu_0}{T_0}-\frac{p^2}{2 m T_0}\right),
\end{equation}
where $m$ is the mass of the dark matter particles, $T_0$ is the temperature of dark matter today, and $\mu_0$ is the chemical potential today, at $a_0\equiv 1$. The temperature $T_0$ determines the mean kinetic energy of the dark matter particles---and hence their mean square velocity $v_\mathrm{rms}=\sqrt{3T_0/m}$---and the chemical potential determines  the number density of dark matter particles at present.   If dark matter has $g_\mathrm{dm}$ degrees of freedom, the right hand side of equation (\ref{eq:EMT}) should be multiplied by $g_\mathrm{dm}$. Since this amounts to a change in the chemical potential $\mu_0$, which in our approach is a free parameter  anyway, we can set  $g_\mathrm{dm}=1$ without loss of generality. 

Because in many models dark matter is in thermal equilibrium in the early universe, the distribution (\ref{eq:MB})  is fairly generic.  If $u^\mu=\delta^\mu_0/a$ is the four-velocity of a comoving observer, then $E=p_\mu u^\mu=\sqrt{m^2+p^2/a^2}$ is the energy of a particle with four-momentum $p_\mu$ in the rest frame of the observer. Hence, the distribution function 
\begin{equation}\label{eq:thermal}
	f(\tau,p_i)=\frac{1}{(2\pi)^3}\exp\left(\frac{\mu-E(p_i)}{T}\right)
\end{equation}
reduces to the distribution (\ref{eq:MB}) in the non-relativistic limit $T/m\ll 1$, and remains a solution of equation (\ref{eq:Boltzmann background}), provided that the temperature  and chemical potential scale appropriately,
\begin{equation}\label{eq:T scaling}
	T=\frac{T_0}{a^2},\quad
	\mu = m+\frac{\mu_0-m}{a^2}.
\end{equation}
Indeed, in order for the distribution to remain time-independent the dark matter temperature has to be inversely proportional to the scale factor, because in the non-relativistic limit $E(p_i)\approx m^2+p^2/a^2$. With this temperature scaling, and in order to again preserve a time-independent distribution, the chemical potential has to be given by the equation above. 

Note however that a thermal distribution of the form (\ref{eq:thermal}) only solves the collisionless Boltzmann equation (\ref{eq:Boltzmann background}) either in the relativistic or non-relativistic limits, but not in both. Hence, our choice of the distribution (\ref{eq:MB}) is justified if dark matter particles kinetically decoupled while non-relativistic.  This is indeed what happens for instance if dark matter consists of WIMPs. In this case, although dark matter typically decouples chemically from the thermal bath while mildly non-relativistic, at $T/m\approx 1/25$, interactions with standard model particles keep dark matter particles in equilibrium with the thermal bath until much later \cite{Bringmann:2006mu}. In contrast, most treatments of warm dark matter models assume that dark matter consists of fermionic particles  which kinetically decouple while highly relativistic, and thus assume that the distribution function follows a  (non-Gaussian) Fermi distribution  with vanishing chemical potential, $f(p)=[1+\exp(p/T_0)]^{-1}$.

Since in the $\Lambda$CDM cosmological  model dark matter particles are assumed to be cold, we usually calculate their energy-momentum tensor in the strict non-relativistic limit ${T/m\to 0}$, in which their energy density  $\rho\equiv -T^0{}_0$ becomes
\begin{equation}
	\bar{\rho}\equiv \frac{\bar{\rho}_0}{a^3}\equiv m \exp\left(\frac{\mu_0-m}{T_0}\right)\left(\frac{m T}{2\pi}\right)^{3/2}.
\end{equation}
 Here, we go beyond this non-relativistic limit and calculate the energy-momentum tensor and its perturbations  to first order in $T/m$. Inserting equation (\ref{eq:MB}) into  (\ref{eq:EMT}) we find
\begin{equation}\label{eq:NR limit}
	\rho=\bar{\rho}\left(1+\frac{3}{2}\frac{T}{m}+\cdots \right),
\end{equation}
and, similarly, the pressure becomes
\begin{equation}
	P\equiv \frac{1}{3}T^i{}_i=\bar{\rho}\frac{T}{m}+\cdots.
\end{equation}
These expressions capture just what we expect from a gas of non-relativistic particles in an expanding universe. Note that $T^0{}_i$ vanishes and $T^i{}_j$ is diagonal, both because of rotational invariance. The  correction factors to the conventional results arise from the thermal average of $p^2$, which decays as $1/a^2$ because physical momenta redshift with the scale factor. It is easy to check that the energy density and pressure above satisfy the conservation equation ${\rho'+3\mathcal{H}(\rho+P)=0}$. The equation of state parameter of this fluid is 
\begin{equation}
	w\equiv \frac{P}{\rho}\approx \frac{T}{m},
\end{equation}
which decays with the square of the scale factor, and is proportional to the small parameter $T/m$. Some authors have constrained the equation of state parameter $w$ of dark matter \cite{Muller:2004yb,Calabrese:2009zza, Serra:2011jh}, but they typically assume that $w$ is a constant, rather than proportional to $1/a^2$.

The expressions above are valid  only at late times, in the non-relativistic limit.  As we proceed back in time the momenta of the dark matter particles increase, and thus become relativistic. As long as dark matter particles remain collisionless, the distribution function (\ref{eq:MB}) remains a solution of the Boltzmann equation (\ref{eq:Boltzmann background}). Therefore, substitution of equation (\ref{eq:MB}) into equation (\ref{eq:EMT}) leads, to all orders in $T/m$, to the energy density
\begin{equation}\label{eq:exact DM density}
	\rho=\bar{\rho}\, \left(\frac{m}{4T}\right)^{1/2}\exp \left(\frac{m}{4T}\right)
		K_1\left(\frac{m}{4T}\right),
\end{equation}
where $K_1$ is  the corresponding modified Bessel function of the second kind and $T$ scales as in equation (\ref{eq:T scaling}). In the limit $m/T\to0$ the energy density scales like that of relativistic particles, whereas in the limit $m/T\to \infty,$ the energy density approaches the limit (\ref{eq:NR limit}).

\subsection{Perturbations}

Our next goal is to derive an equation that captures the impact of a non-zero temperature on the evolution of the dark matter density perturbations.  Under different assumptions and approximations, such an equation has been derived many times in the literature, which extends as far back as to the pioneering work of Gilbert \cite{1966ApJ...144..233G}, the author after whom the equation is mostly named. Most of the relatively  recent derivations of the Gilbert equation have focused on warm dark matter particles, which decouple while relativistic (see e.g. \cite{Bond:1983hb, Brandenberger:1987kf}), although a few analyses have also considered particles that decouple while non-relativistic (see e.g. \cite{Boyanovsky:2008he}). Our derivation here relies on the linearized and relativistic Boltzmann equation in synchronous gauge, and does not involve any approximations beyond an expansion in  $\sqrt{T/m}$, a small parameter in the non-relativistic limit. Our Gilbert equation can thus be incorporated directly into existing numerical Boltzmann codes to calculate CMB anisotropy and matter power spectra in the linear regime. 

In order to study the evolution of structure when dark matter has a non-zero temperature, 
we  perturb the homogeneous and isotropic FRW metric (\ref{eq:FRW}),
\begin{equation}\label{eq:Friedmann perturbed}
	ds^2=a^2\left[-d\tau^2+(\delta_{ij}+h_{ij})dx^i dx^j\right],
	\quad
	h_{ij}=\frac{k_i k_j}{k^2} h+6 \left(\frac{k_i k_j}{k^2}-\frac{1}{3} \delta_{ij}\right)\eta,
\end{equation}
in which we have chosen synchronous gauge, and we concentrate on the perturbation caused by a single Fourier mode of wave vector $\vec{k}$. To calculate the perturbed energy momentum tensor, we need to perturb the background distribution (\ref{eq:thermal}). 
Let us write the perturbed distribution function as
\begin{equation}\label{eq:decomposition}
f(\tau, \vec{x}, \vec{p})=\bar{f}(p)+\delta f(\tau, \vec{x}, \vec{p}),
\end{equation}
where $\bar{f}$ is the thermal distribution (\ref{eq:thermal}) and $p$ is the magnitude of the spatial momentum,
\begin{equation}
	p\equiv a\sqrt{g^{ij} p_i p_j}=\sqrt{\delta^{ij} p_i p_j}-\frac{1}{2}\frac{h_{ij} p_i p_j}{\sqrt{\delta^{ij} p_i p_j}}.
\end{equation}
Note that $p$ depends on the metric, so $\bar{f}$ also contributes to the perturbations of the distribution function. Then, the perturbation $\delta f$ obeys the linearized Boltzmann equation 
\begin{equation}\label{eq:linearized Boltzmann}
	\frac{\partial \delta f}{\partial\tau}+\frac{\partial \delta f}{\partial x^i}\frac{1}{a^2}\frac{p_i}{p^0}-\frac{1}{2}\frac{\bar{f}'}{p} 
	\frac{\partial h_{jk}}{\partial \tau} p_j p_k=0,
\end{equation}
where a prime denotes a derivative with respect to $p$ in this case, and Einstein's summation convention is implied even if repeated indices are not in  opposite locations. It is then easy to check that the perturbed Boltzmann equation admits the line-of-sight integral solution
\begin{multline}\label{eq:delta f}
	\delta f(\tau, \vec{k},\vec{p})=\delta f(\tau_\mathrm{dec},\vec{k},\vec{p})
		\exp\left[-i \, \Delta(\tau,\tau_\mathrm{dec}) \vec{p}\cdot\vec{k}\right]+\\
		{}+ \frac{1}{2}\frac{\bar{f}'}{p}\int_{\tau_\mathrm{dec}}^\tau d\tau' \, 
	p_i p_j \, h_{ij}'(\tau',\vec{k}) 
	\exp\left[-i\,  \Delta(\tau,\tau') \, \vec{p} \cdot \vec{k} \right],
\end{multline}
where
\begin{equation}\label{eq:Delta def}
	\Delta(\tau_2,\tau_1)\equiv \int_{\tau_1}^{\tau_2} d\tau  \frac{1}{a^2(\tau)\, p^0(\tau)} \quad
	\textrm{and}
	\quad
	p^0=\frac{1}{a}\sqrt{m^2+\frac{p^2}{a^2}}.
\end{equation}
Note that $\Delta(\tau_2,\tau_1)\cdot p$  is  the comoving distance traveled by a dark matter particle with covariant momentum $p$ between times $\tau_1$ and $\tau_2$. In particular, $p/(a^2 p^0)$ is its comoving velocity, which, in our non-relativistic approximation can be taken to be $p/(m a)$.  For a thermal distribution, the magnitude of the root mean square momentum is of order $\sqrt{m T_0}$, which leads us to define the comoving free streaming length 
\begin{equation}\label{eq:d def}
	d(\tau_2, \tau_1)\equiv \sqrt{\frac{T_0}{m}}\int_{\tau_1}^{\tau_2} \frac{d\tau}{a}.
\end{equation}
As we shall see shortly, the free streaming captured by the solution (\ref{eq:delta f})   leads to an exponential suppression of structure on comoving scales $k\, d\gg 1$.  

In a universe dominated by matter and radiation,
\begin{equation}\label{eq:explicit d}
	d(\tau_2,\tau_1)=\frac{1}{2}\sqrt{\frac{T_\mathrm{eq}}{m}}\, \tau_\mathrm{eq}
	\log\left(\frac{\tau_2}{\tau_1}\frac{\tau_1+2\tau_\mathrm{eq}}{\tau_2+2\tau_\mathrm{eq}}\right),
\end{equation}
where $T_\mathrm{eq}$ and $(\sqrt{2}-1)\tau_\mathrm{eq}$ respectively are the dark matter temperature and conformal time at matter-radiation equality.  Note that during radiation domination, the product $ \tau \sqrt{T/m}$ is constant, and thus roughly agrees with its value at matter-radiation equality.

We obtain the perturbed energy momentum tensor $\delta T^\mu{}_\nu$ by substituting the solution (\ref{eq:delta f}) into equation (\ref{eq:EMT}). As we describe in detail in Appendix B, the perturbed energy density becomes
\begin{equation}\label{eq:delta rho}
\begin{split}
	\delta\rho=
	-\frac{\bar\rho}{2}\int_{\tau_\mathrm{dec}}^\tau d\tau' e^{-d^2 k^2/2}
	\Big\{&h'-(d\,  k)^2 (h'+4\eta')+{}\\
	&{}+\frac{T(\tau)}{2m}
	\left[(5- d^2 k^2)h'- (d\, k)^2 (7- d^2\, k^2) (h'+4\eta')\right]
	\Big\}.
\end{split}
\end{equation}
In this equation, the free streaming length $d=d(\tau,\tau')$ is given by equation (\ref{eq:d def}) and we have assumed that the perturbation $\delta f$ vanishes at decoupling. Note the exponential factor inside the integrand, which suppresses the contributions of the potentials  on scales ${(d\, k)^2 \gg 1}$. The exponential arises from the moments of spherical Bessel functions with respect to the Gaussian distribution in equation (\ref{eq:MB}), and is thus sensitive to the precise form of the distribution function.  Because the comoving free streaming length $d$ depends on the dark matter temperature, the absence of such suppression allows us to place quite stringent constraints on $T/m$.  The CDM density contrast
\begin{equation}
	\frac{\delta\rho}{\rho}\approx \frac{\delta\rho}{\bar{\rho}}\left(1-\frac{3}{2}\frac{T}{m}\right)
\end{equation}
contains an additional  correction due to the non-zero dark matter temperature, but the impact of this correction is typically much smaller than that due to the free-streaming term. 

As opposed to what happens  in the limit $T/m\to 0$, in which the dark matter velocity can be taken to vanish in synchronous gauge, in this case the velocity potential is non-zero.  With $\delta T^0{}_i\equiv (\rho+P) \partial_i v$, we find 
\begin{equation}\label{eq:v}
	(\rho+P) v=\frac{\bar\rho}{2}\sqrt{\frac{T}{m}} \frac{1}{k}
	\int_{\tau_\mathrm{dec}}^\tau  d\tau' e^{-d^2 k^2/2} 
	 \left[  d\, k \, (3h'+8\eta') - (d\, k)^3 (h'+4\eta')\right],
\end{equation}
which shows an analogous suppression of the velocity perturbation on scales much smaller than $d$. Finally, following the same approach, we arrive at 
\begin{displaymath}
\begin{split}
	\delta T^i{}_j=-\frac{\bar\rho}{2}\frac{T}{m}\int_{\tau_\mathrm{dec}}^\tau d\tau'
	e^{-d^2 k^2/2}
	\Big\{ & h'\delta_{ij}+2h'_{ij}+ (d\, k)^2\left[(h'+4\eta')\delta_{ij}
	+(5h'+16\eta')\hat{k}_i\hat{k}_j\right]+{} \\
	& {}+ (d\, k)^4 (h'+4\eta') \hat{k}_i\hat{k}_j
	\Big\},
\end{split}
\end{displaymath}
from which we can immediately read off the perturbed pressure $\delta p$ and the scalar anisotropic stress $\pi$,  $\delta T^i{}_j\equiv \delta p \, \delta^i_j-k^i k_j \, \pi$,
\begin{align}
	\delta P&=-\frac{\bar\rho}{2} \frac{T}{m} \int_{\tau_\mathrm{dec}}^\tau d\tau' \, 
	e^{-d^2 k^2/2} \left[h'-4\eta' + (d\, k)^2 (h'+4\eta')\right], \label{eq:delta p}\\
	\pi &=\frac{\bar\rho}{2} \frac{T}{m} \frac{1}{k^2}\int_{\tau_\mathrm{dec}}^\tau d\tau' \, 
	e^{-d^2 k^2/2} \left[2h'+12\eta' + (d\, k)^2 (5h'+16\eta') + (d\, k)^4 (h'+4\eta')\right].
	\label{eq:pi}
\end{align}
Note that the contribution of the anisotropic stress to the energy momentum tensor, of order $k^2 \pi$, is of the same magnitude as that of the pressure perturbation. The pressure perturbation itself is a factor $T/m$ smaller than the energy density perturbation, as expected. 
 
With $\delta\rho$, $\delta P$, $\pi$ and $\delta u$ given by the previous expressions, it is relatively straightforward, albeit tedious, to verify that the energy momentum tensor is covariantly conserved up to terms of order $(T/m)^{3/2}$. In particular, these quantities obey the perturbed hydrodynamical equations
of energy conservation,
\begin{equation}
	\delta\rho'+3\mathcal{H}(\delta\rho+\delta P)-k^2 \left[(\rho+P) v+\mathcal{H}\pi\right]+(\rho+P)\frac{h'}{2}=0,
\end{equation}
and momentum conservation,
\begin{equation}
	[(\rho+P) v]'+4\mathcal{H} (\rho+P) v+\delta P-k^2\pi=0.
\end{equation}
In any case, because the anisotropic stress is of the same order as the pressure perturbation, a (perfect) fluid description of dark matter breaks down on small scales. For instance, as we mention in Appendix \ref{sec:Calculation},  collisionless dark matter does not undergo acoustic oscillations, even if the dark matter particles have a non-zero velocity dispersion, and thus, a non-zero pressure. 

\section{Impact on Structure Formation}
\label{sec:Impact}
Here, we are interested in assessing the impact of a non-zero CDM temperature on the formation of  structure at  scales accessible to linear perturbation theory. At present, constraints on the linear power spectrum at these smallest scales rely on  the Lyman-alpha forest \cite{McDonald:2004eu}, which  probes  comoving wave numbers of order
 \begin{equation}\label{eq:k max}
 	 k_\mathrm{max} \approx 2\ h\,  \mathrm{Mpc}^{-1}
 \end{equation}
at redshifts $z\approx 3$. This should be compared with the wavenumber we can probe with the $\ell$'th multipole of the  cosmic microwave background, 
\begin{equation}\label{eq:k CMB}
	k_\mathrm{CMB}\approx 0.21 \frac{\ell}{3500} \mathrm{Mpc}^{-1},
\end{equation}
which is about an order of magnitude smaller than $k_\mathrm{max}$ even for the angular scales probed by ACT \cite{Das:2013zf}, SPT \cite{Story:2012wx} and Planck \cite{Planck:2013kta}.

In a $\Lambda$CDM cosmology the scale $k_\mathrm{max}$ enters the horizon at a redshift of about ${z_\mathrm{max}\approx 5\cdot 10^5}$.  Because our analysis assumes that at redshift $z_\mathrm{max}$ cold dark matter particles were already  non-relativistic, the temperature today hence needs to obey
\begin{equation}\label{eq:validity}
	\frac{T_0}{m}\lesssim 2\cdot 10^{-12},
\end{equation}
and has to be proportionally smaller if we are interested in length scales smaller than $k_\mathrm{max}$. In addition, because we also assume that dark matter is collisionless, it needs to decouple at redshift $z_\mathrm{dec}>z_\mathrm{max}$.  This behavior should be contrasted with that of neutrinos or warm dark matter, which decouple kinetically  while being relativistic and become non-relativistic after galactic scales have entered the horizon. The impact of a non-zero dark matter temperature on scales with wave numbers much larger than (\ref{eq:k max}), as well as the imprint  of dark matter decoupling on the matter power spectrum, is discussed in \cite{Green:2005fa,Loeb:2005pm}.
 
Our next goal is to estimate the matter  density perturbation (\ref{eq:delta rho}) after matter-radiation equality, at $\tau_0>\tau_\mathrm{eq}$. At recombination, the density of dark matter has an impact on the structure of the CMB Doppler peaks, and at redshift zero, the dark matter perturbation is directly related to the matter power spectrum.  Because dark matter particles are non-relativistic, we assume that we are in a regime in which we can drop the term proportional to $T/2m$ in equation (\ref{eq:delta rho}), which in this approximation becomes
\begin{equation}\label{eq:split}
	\delta\rho=-\frac{\bar\rho(\tau_0)}{2}\left\{\int_{\theta_\mathrm{dec}}^{\theta_\mathrm{eq}}
	 d\theta
	+\int_{\theta_\mathrm{eq}}^{\theta_0} d\theta \right\}
	e^{-d^2 k^2/2}\left[\frac{dh}{d\theta}-(d\,  k)^2
	 \left(\frac{dh}{d\theta}+4\frac{d\eta}{d\theta}\right)\right].
\end{equation}
Note that we have split the integral into the contribution to $\delta\rho$ during radiation domination and that during matter domination, and that we have introduced the new integration variable
 \begin{equation}
 	\theta=\frac{k\tau}{\sqrt{3}}.
 \end{equation}

 \subsection{Radiation domination}
Well in the radiation-dominated era, the free-streaming length in equation (\ref{eq:explicit d})  becomes
 \begin{equation}\label{eq:explicit d rad}
 d(\theta_0, \theta)\approx  \frac{1}{2}\sqrt{\frac{T_\mathrm{eq}}{m}} \tau_\mathrm{eq}
	\log\left(2\frac{\theta_\mathrm{eq}}{\theta}\right), \quad \quad \theta\ll \theta_\mathrm{eq},
 \end{equation}
During this era, we can neglect the impact of dark matter on the gravitational potentials, which take their standard values
 \begin{equation}\label{eq:derivatives}
	 \frac{dh}{d\theta}=\frac{12\mathcal{R}_i}{\theta}
	 \left(\frac{2(\cos \theta-1)}{\theta^2}+\frac{2 \sin \theta}{\theta}-1\right),\quad
 	\frac{d\eta}{d\theta}=-\frac{4\mathcal{R}_i}{\theta}
	\left(\frac{\sin \theta}{2\theta}-\frac{1-\cos \theta}{\theta^2}\right),
 \end{equation}
where $\mathcal{R}_i$ is the initial (primordial) curvature perturbation.

There are two dimensionless ratios that determine the behavior of the perturbations: $\theta_\mathrm{eq}$ and the ratio of wavelength to the free-streaming length in equation (\ref{eq:explicit d rad}), $k\, d$. In order to proceed, we focus on the short-wavelength limit $\theta_\mathrm{eq}\gg 1$ and discuss  the limits $k\,d\ll 1$ and $k\, d\gg 1$, in this order, separately. 

In the absence of free-streaming ($k\, d\equiv 0$), the dominant contributions to the integral in equation (\ref{eq:split}) stem from the extrema of $dh/d\theta$ and $d\eta/d\theta$  at $\theta_\mathrm{max}\approx 3$. Because the logarithmic derivative of $(k \, d)^2$ is
\begin{equation}
	\frac{d\log (k\, d)^2}{d\theta}=-\frac{2}{\theta}\log^{-1} \frac{2\theta_\mathrm{eq}}{\theta},
\end{equation}
in the limit  of large $\theta_\mathrm{eq}$  both the exponential and the factor of  $(kd)^2$ in the integrand are then slowly varying functions of $\theta$ around $\theta_\mathrm{max}$ for small $k\, d$.  Therefore, taking the factors of $k\,d$ out of the integral, and evaluating at the extrema of the corresponding potential derivatives we get 
\begin{equation}\label{eq:linear effect}
	\frac{\delta\rho}{\bar{\rho}}\approx 6\mathcal{R}_i\left(\log \theta_\mathrm{eq}+\gamma-\frac{1}{2}\right)-\frac{\mathcal{R}_i}{8}\,\frac{T_\mathrm{eq}}{m}  \, \theta_\mathrm{eq}^2 \log^2 \frac{2\theta_\mathrm{eq}}{3}  \times\left(18\log\theta_\mathrm{eq}+18\gamma-13\right),
\end{equation}
where $\gamma$ is Euler's constant. As expected, free streaming suppresses density perturbations at small scales, although, by assumption,  the effect is small in this limit.

When $k\, d$ is large, the exponential in equation (\ref{eq:split}) is a rapidly varying function of $\theta$, which strongly suppresses the contribution of a mode when $\theta\ll \theta_\mathrm{eq}$. Therefore, in this limit, the integral is dominated by the values of the integrand around $\theta=\theta_\mathrm{eq}$. Assuming constant derivatives of the gravitational potentials around that point, and taking those functions outside the integral we thus get
\begin{equation}
	\frac{\delta\rho}{\bar{\rho}}\approx -6 \, \mathcal{R}_i \log 2 
	\exp\left[-\frac{3}{8}(T_\mathrm{eq}/m) \theta_\mathrm{eq}^2 \log^2 2\right],
\end{equation}
 where we have only kept the contribution of the term proportional to $(d\, k)^2$. In this case, the exponential suppression of the density perturbations essentially smoothes out  dark matter inhomogeneities  on comoving scales with $\theta_\mathrm{eq}^2\gg m/T_\mathrm{eq}$.
 
\subsection{Matter domination}
It still remains to calculate the contributions to (\ref{eq:split}) from the matter-dominated era. During this epoch,  the comoving free-streaming length in equation (\ref{eq:explicit d}) becomes
\begin{equation}\label{eq:explicit d mat}
	d(\theta_0, \theta)\approx  \frac{1}{2}\sqrt{\frac{T_\mathrm{eq}}{m}} \tau_\mathrm{eq}
	\log\left(1+2\frac{\theta_\mathrm{eq}}{\theta}\right), \quad \quad \theta\gg \theta_\mathrm{eq}.
\end{equation}
Hence, well in the matter-dominated era, the free-streaming length is suppressed by a factor of order  $\theta_\mathrm{eq}/\theta$ relative to that of the streaming length during radiation domination, even though most of the growth in $\delta\rho/\rho$ happens during this time. As a result, in the limit in which free-streaming has a sizable impact on structure formation, we expect that impact to be largest during the radiation-dominated regime.  

\begin{figure}[t!]
\begin{center}
\includegraphics{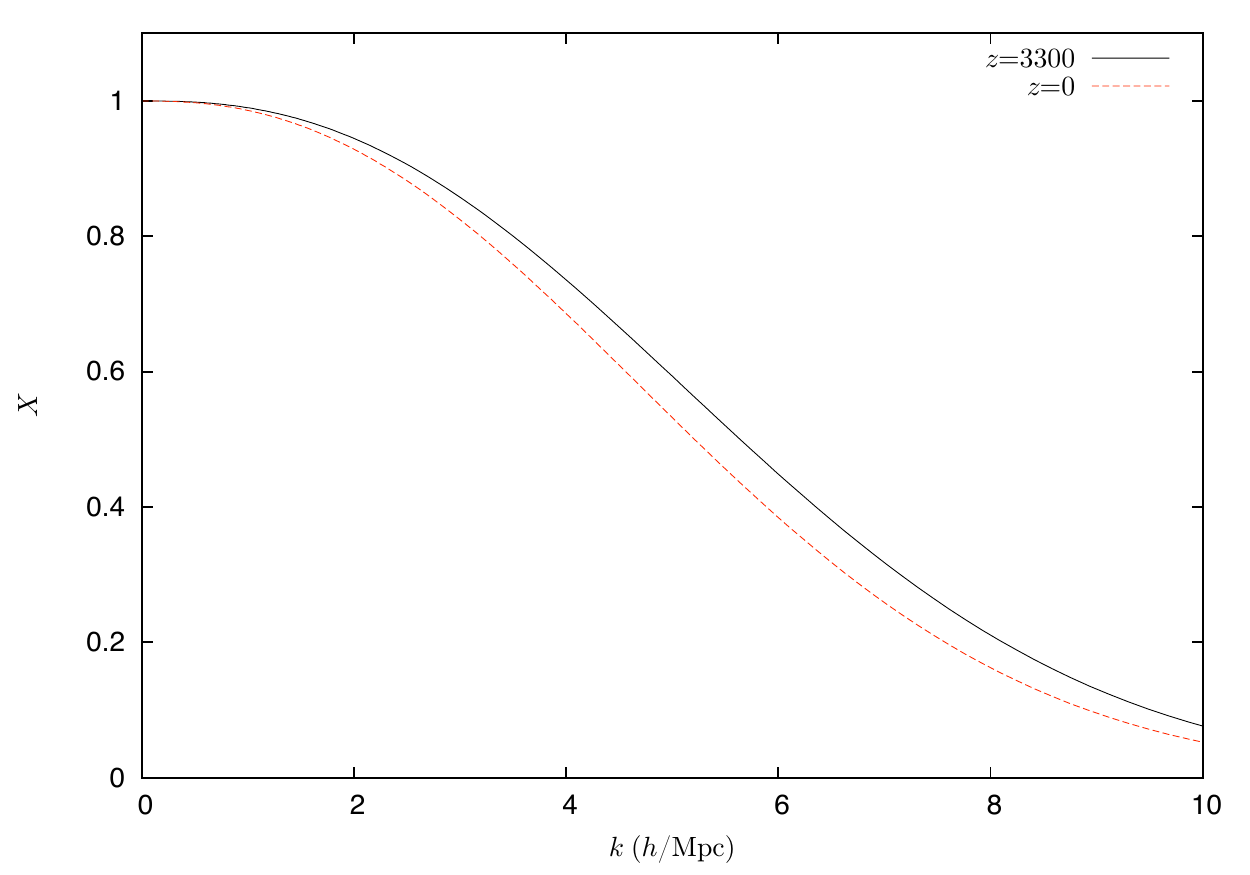}
\end{center}
\caption{A plot of the dark matter transfer function ratio $X$ in equation (\ref{eq:R}) for fixed temperature $T_0/m=10^{-7}$ at redshifts $z=3300$ (black continuous) and $z=0$ (red dashed).  Most of the impact of free streaming on the suppression of dark matter inhomogeneities occurs before matter-radiation equality, at $z\approx 3300$.} \label{fig:ratios}
\end{figure}

To illustrate the impact of free-streaming during the matter-dominated epoch, consider for instance the ratio 
\begin{equation}\label{eq:R}
	X\equiv \frac{\mathcal{T}(k,z,T_0/m)}{\mathcal{T}(k,z,0)},
\end{equation}
where $\mathcal{T}(k,z,T_0/m)$ is the dark matter transfer function at wave number $k$, redshift $z$ and present dark matter temperature $T_0/m$.  In Figure  \ref{fig:ratios},  we plot  $X$ as a function of $k$ for fixed $T_0/m=10^{-7}$ at redshifts $z=3300$ and $z=0$.  As seen in the figure, at $z=3300$ most of the suppression of dark matter inhomogeneities (due to free streaming) is already in place. There is an additional suppression at $z=0$, but the latter is not significantly different from that at $z=3300.$

Equations (\ref{eq:explicit d rad}) and (\ref{eq:explicit d mat}) allow us to obtain a rough estimate of the comoving length scale below which free streaming leads to a suppression of dark matter anisotropies. Both equations show that the free streaming length $d$ responsible for the exponential suppression of structure in equation (\ref{eq:split}) equals, modulo a logarithmic factor, $\sqrt{T_\mathrm{eq}/m}\, \tau_\mathrm{eq}/2$.  We are thus led to define the comoving free-streaming wave number
\begin{equation}\label{eq:kfs}
	k_\mathrm{fs}=\sqrt{\frac{m}{T_\mathrm{eq}}}\frac{2}{\tau_\mathrm{eq}}\approx \sqrt{\frac{m}{T_\mathrm{eq}}} \frac{\Omega_m h^2 }{16\, \mathrm{Mpc}}.
\end{equation}
Noting that the dark matter temperature today $T_0$ is related to that at equality by
\begin{equation}
	T_\mathrm{eq}= (1+z_\mathrm{eq})^2 T_0 \approx 
		(2.3 \cdot 10^4\,  \Omega_m h^2)^2 \, T_0,
\end{equation}
we thus obtain the free-streaming wave number
\begin{equation} \
	k_\mathrm{fs}\approx 2.6\cdot 10^{-6} \sqrt{\frac{m}{T_0}} \, \mathrm{Mpc}^{-1},
\end{equation}
which does not depend on the dark matter density.  This is basically the length scale derived in references \cite{Bond:1983hb, Bode:2000gq}, although the scalings with $\Omega_m h^2$ differ. The reason is that whereas in our analysis $T_0/m$  and $\Omega_m h^2$ are independent parameters, in warm dark matter models, the current dark matter density and the dark matter mass are used to determine the dark matter temperature (we are ignoring baryons here). Modulo the logarithmic factor that we dropped, the free-streaming scale (\ref{eq:kfs}) also agrees with the scale  derived for WIMPs in \cite{Green:2005fa}, provided that equations with the same parameters are compared.  For further analytical estimates of the impact of a non-zero temperature on the matter power spectrum during matter domination, see reference \cite{Boyanovsky:2008he}.

\subsection{Power Spectra}

We have seen that a non-vanishing dark matter temperature generically leads to a suppression of structure on small scales. In order to determine the quantitatively precise nature of this suppression, we need to rely on a numerical solution of the perturbation equations.

\begin{figure}[t!]
\begin{center}
	\includegraphics{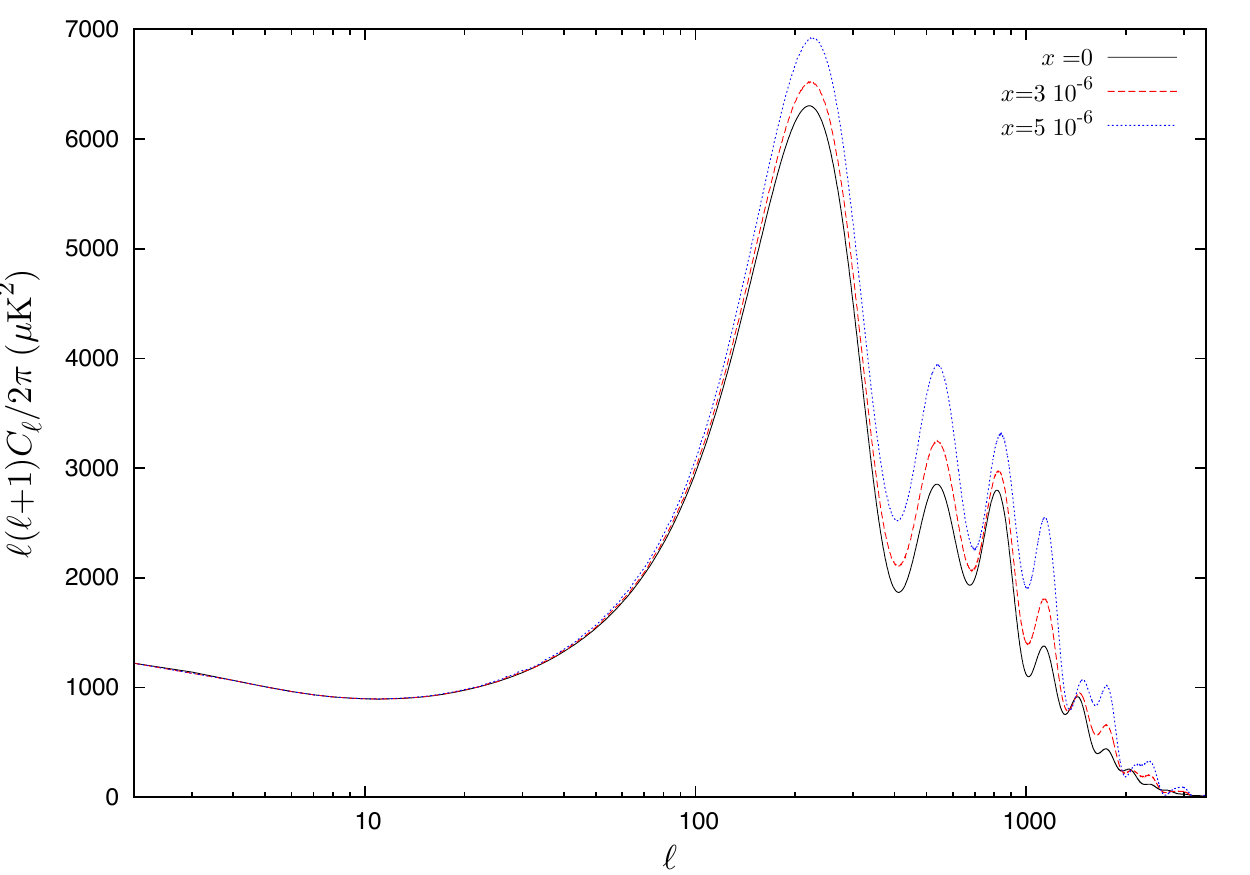}
\end{center}
\caption{Temperature anisotropy  power spectra for different values of $x\equiv \sqrt{T_0/m}$  as a function of spherical multipole $\ell$.  Differences in the angular power spectra with respect to the $x=0$ case have been magnified by a factor of $10^2$. } \label{fig:Cls}
\end{figure}

In Figure \ref{fig:Cls}, we plot the temperature anisotropy power spectrum  for different values of the dark matter temperature. Although these temperatures do not satisfy the condition (\ref{eq:validity}), they are low enough for our non-relativistic approximation to be trusted, since, according to equation (\ref{eq:k CMB})  the comoving scales probed by the CMB are smaller than $k_\mathrm{max}$ in equation (\ref{eq:k max}).  Even at these temperatures, the impact of a non-zero temperature on the CMB power spectrum is not visible, so we have magnified the difference with respect to the the $T=0$ power spectrum by a factor of a hundred.  As seen in the figure, a non-zero temperature does not shift the location of the acoustics peaks significantly, which simply grow in amplitude,  the  growth being more pronounced for the even peaks.  At least partially, this behavior is relatively simple to explain: in synchronous gauge, and in the approximation of instantaneous recombination, the temperature anisotropies are mostly determined by the Sachs-Wolfe term  at last scattering \cite{Weinberg:2008zzc}
\begin{equation}\label{eq:SW}
	F(k)=\frac{\mathcal{R}_i}{5}
	\left[3\mathcal{T}(k/k_\mathrm{eq})R_L-\frac{\mathcal{S}(k/k_\mathrm{eq})}{(1+R_L)^{1/4}}
	\exp\left(-\int_0^{\tau_L} \Gamma d\tau\right)
	\cos\left(k \int_0^{\tau_L} \frac{d\tau}{\sqrt{1+R}}\right)\right],
\end{equation}
where $\sqrt{2}k_\mathrm{eq}$ is the mode that enters the horizon at recombination, $R$ is the baryon to photon density ratio, $\mathcal{T}$ is the dark matter transfer function, and $\mathcal{S}$ is the  transfer function that determines the amplitude of the photon acoustic oscillations.   A non-zero dark matter temperature hardly impacts the amplitude of the acoustic oscillations $\mathcal{S}$,  since the latter is determined during radiation domination, but it does suppress the transfer function $\mathcal{T},$ because free-streaming damps dark matter perturbations. As a result, the source term $F$ increases in magnitude at the location of the even peaks (where the cosine in equation (\ref{eq:SW}) is positive), leading to a power increase in the even peaks of the angular power spectrum. Of course, by the same token we would expect a power decrease at the odd peaks; instead we just observe a less prominent increase in the peak amplitude.

\begin{figure}[t!]
\begin{center}
	\includegraphics{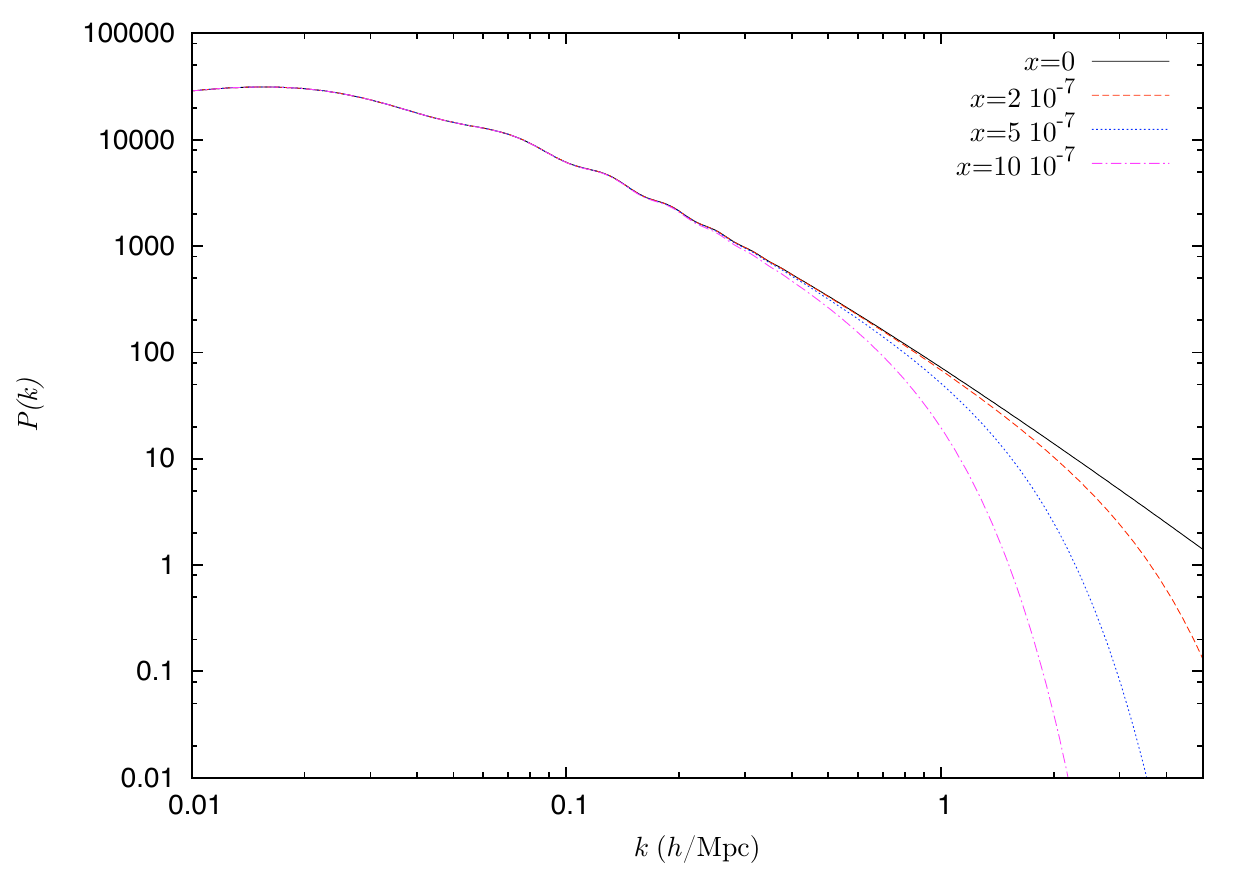}
\end{center}
\caption{Matter   power spectrum for different values of $x\equiv \sqrt{T_0/m}$  as a function of comoving wave number $k\, h\, \mathrm{Mpc}$. On length scales smaller than the free streaming length, structure is suppressed. Because the free-streaming length is proportional to the dark matter temperature, larger temperatures lead to more suppression.} \label{fig:matterpower}
\end{figure}

The effects of a non-zero dark matter temperature are more pronounced in the (total) matter power spectrum at $z=0$, which is a far more direct probe of the dark matter distribution. In Figure \ref{fig:matterpower}, we  plot the matter power spectrum also for a different set of dark matter temperatures. Whereas the impact of a non-zero temperature on the CMB was hardly visible, here,  departures in the matter power spectrum are very prominent   on scales $k>0.4\,  h\,  \mathrm{Mpc}^{-1}$, and show the expected suppression due to the free streaming of dark matter particles. Of course at these scales linear perturbation theory breaks down, so our linear calculation has to be appropriately interpreted.\footnote{If  the dark matter temperature is high enough, the associated suppression of structure may keep all scales in the linear regime. Obviously, there would not be any collapsed haloes in such a universe, which would be very different from ours. } 

Ma  has found that in warm dark matter models the ratio of dark matter linear transfer functions $X$ in equation (\ref{eq:R}) is well fit by
\begin{equation}\label{eq:fit}
	X\approx\frac{1}{\left[1+(\alpha k)^{2\nu}\right]^{5/\nu}},
\end{equation}
where $\alpha$ depends on various cosmological parameters, such as $\Omega_m$ and the mass of the warm dark matter particle,  and the exponent $\nu$ is  a constant, $\nu\approx 1.2$ \cite{Bode:2000gq}. Although our non-relativistic approximation does not allow us to numerically explore the regime $k\, d\gg 1$ in which we expect structure to be exponentially suppressed, we find that different exponents $\nu$ provide better fits to the numerical results as we vary the  dark matter temperature. Say, for $T_0/m=10^{-16}$ the exponent $\nu\approx 1.17$ gives a squared sum of square residuals about  thirty times smaller than for $\nu=1.2$, whereas for $T_0/m=10^{-14}$, $\nu\approx 1.25$ gives sum of square residuals about three times smaller than for $\nu=1.2$.  

\begin{figure}[t!]
\begin{center}
	\includegraphics{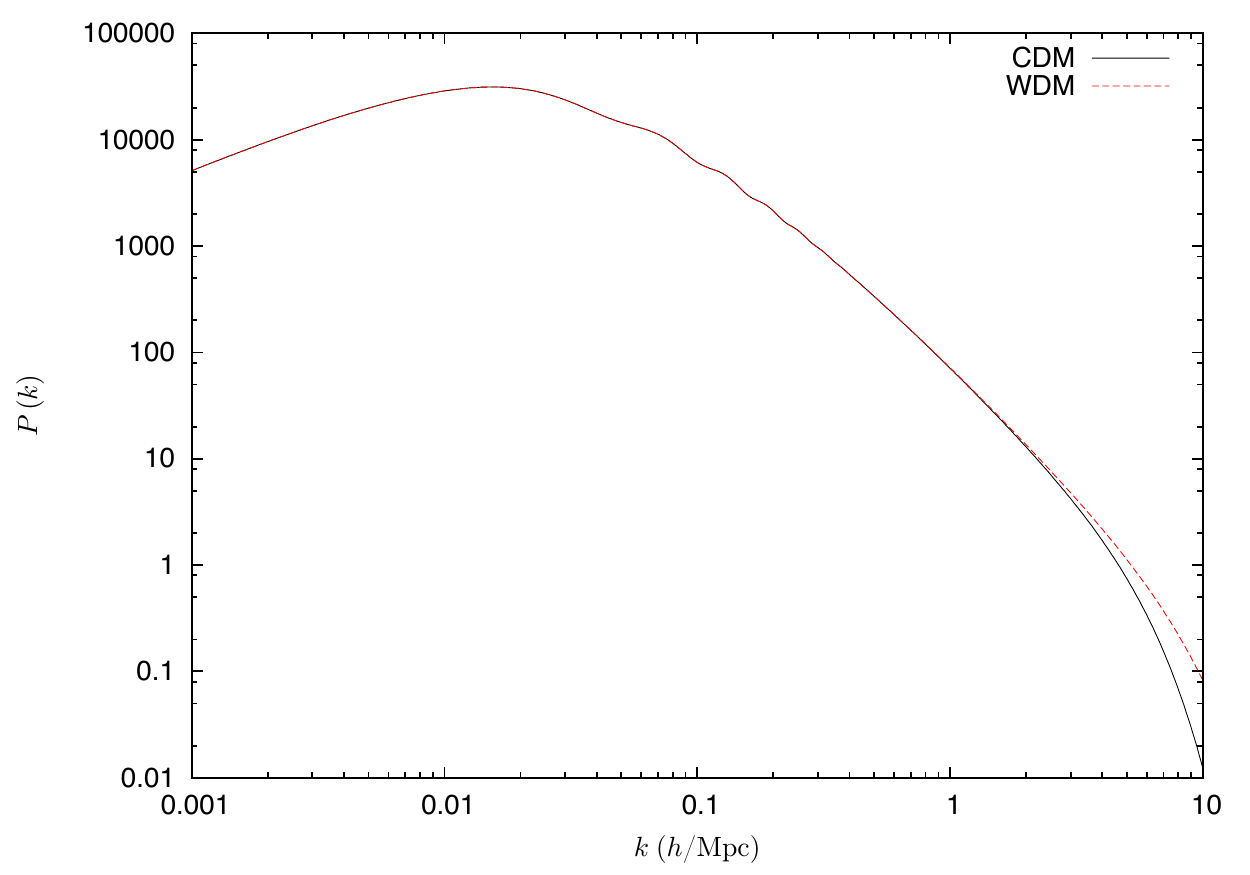}
\end{center}
\caption{Comparison of a CDM matter power spectrum at $T_0/m=10^{-14}$ with a WDM obtained from the fitting formulate in equation (\ref{eq:fit}), with $\nu=1.2$ and the value of $\alpha$ that matches the CDM velocity dispersion at present. The remaining cosmological parameters have the same values.}
 \label{fig:WDM}
\end{figure}

We have not explored however how the parameter $\alpha$ depend on the dark matter temperature or the remaining cosmological parameters. If for a given  non-zero CDM temperature and fixed cosmological parameters we simply determine the value of $\alpha$ in equation (\ref{eq:fit}) that best fits the the cold dark matter spectrum for fixed $\nu=1.2$,  we  find an excellent agreement between both. This agreement between the CDM and WDM spectra is what one would expect from the relative similarity of the exponent $\nu$ described above.  On the other hand, when we compare  CDM and WDM spectra with the same cosmological parameters  and the same velocity dispersion at present\footnote{We fix the mass of the warm dark matter particle by matching  equation (A3) of reference \cite{Bode:2000gq} to the desired CDM velocity dispersion $v_\mathrm{rms}=\sqrt{3T_0/m}$. The resulting mass is then substituted into equation (A9) of \cite{Bode:2000gq},  which then determines the coefficient $\alpha$.} we find a significant disagreement at small scales, as shown in Figure \ref{fig:WDM} .

\section{Limits}
The  results of the previous section allow us to place a rough but conservative limit on the dark matter temperature today. A host of cosmological measurements of small scale structure seem to be in good agreement with the standard $\Lambda$CDM cosmological model. As we saw in Section \ref{sec:Impact}, the CMB is not very sensitive to the dark matter temperature on these small scales. On the other hand, the distribution of large scale structure is directly affected by a non-zero dark matter temperature, and can probed  down to $k_\mathrm{max}\approx 2\, h\, \mathrm{Mpc}^{-1}$ by  Lyman-alpha forest observations in  reference \cite{McDonald:2004eu}. Therefore,  we expect the most stringent constraints on the dark matter velocity to arise  from   measurements of the dark matter power spectrum on these scales. 

The most recent (published)  analysis of how the Lyman-alpha forest constrains the matter power spectrum is that of reference  \cite{McDonald:2004eu}. The measurement suffers from significant systematic errors, affecting the amplitude of the power spectrum at $z=3$ and $k=2 \, h\, \mathrm{Mpc}^{-1}$ by factors of up to $25\%$. Demanding then that the relative correction  to the dark matter overdensity in equation (\ref{eq:linear effect}) be less than $50\%$ on those scales we thus arrive at the  limit $T_\mathrm{eq}/m \lesssim 10^{-6}$. Because the temperature is inversely proportional to $a^2$, and $1+z_\mathrm{eq}\approx 3\cdot  10^3$, this implies that $T_0/m\lesssim  10^{-13}$. Since this ratio is  smaller that the one necessary for the validity our approximation, equation (\ref{eq:validity}), our analysis is at the very least self-consistent.   We derive  a sharper numerical limit next.

\subsection*{Numerical Results}
In order to place rigorous and precise marginalized limits on the temperature to mass ratio, we resort to the by-now standard  Bayesian approach to parameter estimation based on Markov-Chain Monte Carlo methods. We have modified the publicly available Boltzmann integrator CAMB and the Markov-Chain Monte Carlo engine CosmoMC \cite{cosmomc, Lewis:1999bs, Lewis:2002ah} by  including the necessary modifications of the dark matter equations needed to account for a non-zero dark matter temperature, as detailed in Appendix \ref{sec:Numerical Implementation}. We sample the posterior probability for a spatially flat cosmological model with parameters $H_0$ (Hubble's constant today), $\Omega_\Lambda$ (critical density fraction of a cosmological constant), $\Omega_b h^2$ (baryon density), $\tau$ (optical depth), $n_s$ (scalar spectral index), $A_s$ (scalar spectral amplitude), $A_\mathrm{SZ}$ (amplitude of a Sunyaev-Zeldovich template)  and $\sqrt{T_0/m}$ (square root of present dark matter temperature to mass ratio)   with a set of four Monte Carlo Markov chains of at least $2 \times 10^5$ elements each, generated with an appropriately modified version of CosmoMC. We impose flat priors on all  parameters, assume that the universe is spatially flat and neglect tensor modes. To check for the converge of our chains, we monitor the Gelman and Rubin statistic \cite{Gelman92}, which stays under $10^{-2}$. Following CosmoMC output, we also estimate the statistical errors on our upper limits by exploring their changes upon split of our chains in several subsamples, which remain of the order of $1\%$.

In order to obtain the strictest constraints on the dark matter temperature, it is crucial to  employ observations at small scales. We thus include constraints on the linear matter power spectrum at redshift $z=3$ derived from Lyman alpha observations in reference \cite{McDonald:2004eu} (surprisingly, this 2005 analysis is still state-of-the-art). Measurements of the linear power spectrum on  the scales probed by the Lyman alpha forest are notoriously difficult, and typically require structure formation simulations for various input power spectra and cosmological parameters. Although the simulations  carried in reference \cite{McDonald:2004eu} just involved the standard $\Lambda$CDM cosmological model, their constraints on the linear power spectrum should remain valid as long as the linear matter power spectrum does not significantly deviate from that in $\Lambda$CDM. We enforce such an agreement at the corresponding scales with the temperature prior
\begin{equation}\label{eq:prior}
	\sqrt{\frac{T_0}{m}}\leq 2\cdot 10^{-7},
\end{equation}
which also guarantees the validity of our perturbative equations.

\begin{figure}[t!]
\begin{center}
	\includegraphics{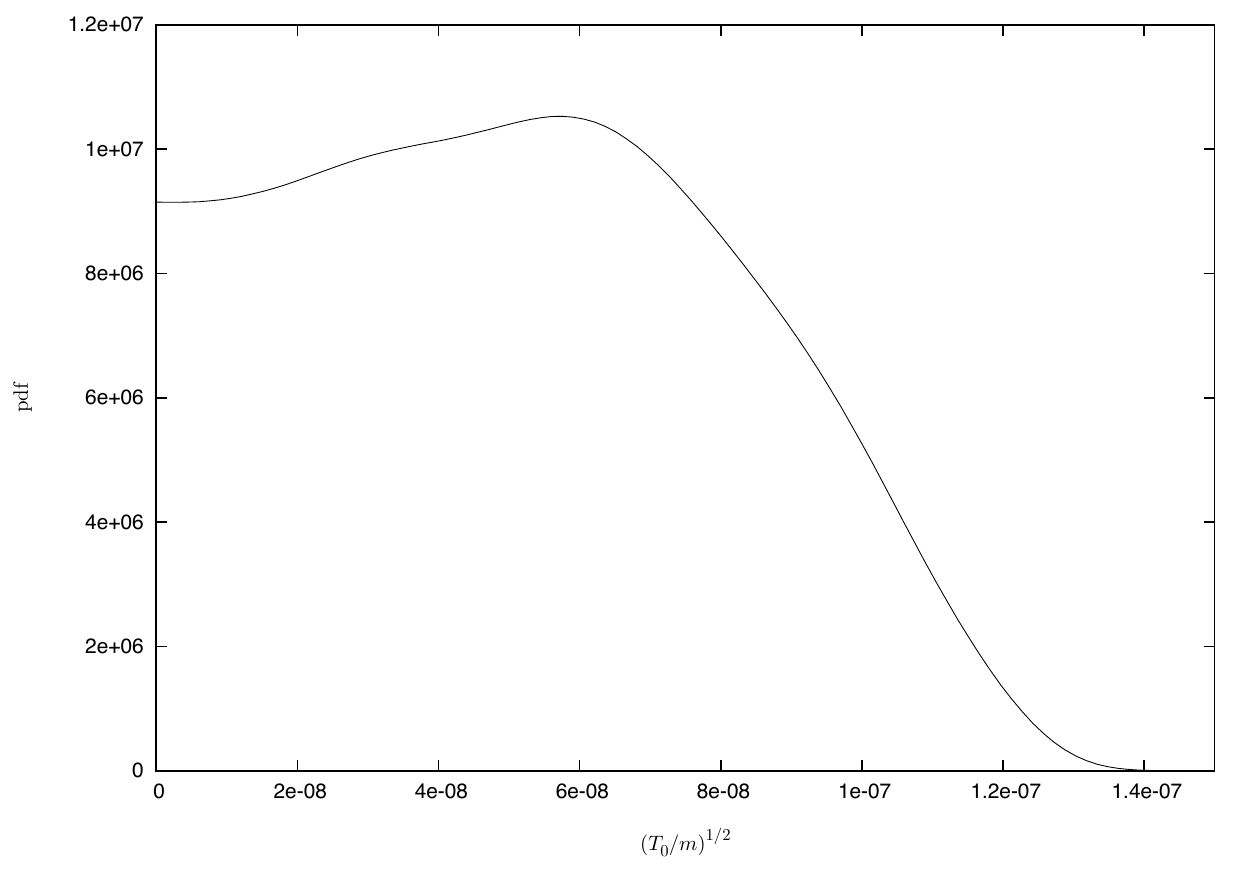}
\end{center}
\caption{Marginalized posterior distribution of $\sqrt{T_0/m}$. Note the relatively flat plateau at low temperatures, which indicates that data cannot discriminate between temperatures in the range $\sqrt{T_0/m}\lesssim 6\cdot 10^{-8}$. \label{fig:posterior}}
\end{figure}

Although at the temperatures of interest the cosmic microwave background is hardly affected,  observations of the cosmic microwave background are nevertheless crucial to constrain the remaining cosmological parameters. We therefore include cosmic microwave measurements from the WMAP 9 year data release \cite{Bennett:2012fp}, as well as ACT  \cite{Das:2013zf} and SPT data \cite{Story:2012wx}, which probe the angular power spectrum on smaller scales (the first year Planck mission power spectrum measurements  \cite{Planck:2013kta} have not been published at the time of this writing). We also include large scale structure data from an SDSS luminous red galaxy (LRG) sample \cite{Reid:2009xm}. We calculate the likelihood of our angular power spectra with the numerical codes supplied by the corresponding collaboration \cite{Dunkley:2013vu}, and we employ the patch\footnote{http://www.slosar.com/aslosar/lya.html} written by  An\v{z}e Slosar to evaluate matter  power spectra likelihoods on Lyman-alpha scales \cite{McDonald:2004eu}.

Proceeding as outlined above,  using  the afore-mentioned datasets, we obtain the marginalized posterior distributions  for $\sqrt{T_0/m}$ shown in Figure \ref{fig:posterior}.  We also list mean, standard deviation and   credible upper limits of the corresponding posterior distribution in Table \ref{table:statistics}. Simple inspection of the posterior distribution shows that there is no evidence for a non-zero dark matter temperature in the data. In fact, adding the temperature to mass ratio $\sqrt{T_0/m}$ to the standard cosmological parameters in $\Lambda$CDM improves the log likelihood just by $0.15$.  The  $95\%$ upper credible limit on the dark matter to temperature ratio then is
\begin{equation}\label{eq:limit}
	\frac{T_0}{m}\leq 1.07 \cdot 10^{-14},
\end{equation}
which translates into an upper limit on the  present dark matter rms velocity $v_\mathrm{rms}\leq 54\, \mathrm{m/s}$. The mean of the posterior distribution of $T_0/m$  is several standard deviations away from the edge of our prior (\ref{eq:prior}), which therefore has no influence on the upper limit (\ref{eq:limit}). These  results do not depend on any particular model, and only rely on the assumptions that dark matter is collisionless, and  that  the distribution of its momenta is Maxwellian, with a temperature $T/m\ll 1$. As we emphasized previously, equation (\ref{eq:limit}) should not be interpreted as a constraint on the actual dark matter temperature at present (because the latter is mostly found in collapsed haloes),  but as an extrapolation: The limit implies that at redshift $z$, where $1\ll z\leq z_\mathrm{dec}$, the dark matter temperature has to obey
\begin{equation}\label{eq:f of z limit}
	\frac{T}{m}\leq  1.07 \cdot 10^{-14} (1+z)^2.
\end{equation}
Imagine, for example, that a dark matter model  predicts a decoupling redshift $z_\mathrm{dec}$, at which dark matter matter decouples kinetically. If this decoupling redshift obeys ${z_\mathrm{dec}>z_\mathrm{max}\approx 5\cdot 10^5}$ the assumptions of our analysis hold. Then, evaluating  the  inequality (\ref{eq:f of z limit}) at $z_\mathrm{dec}$, we  obtain an actual  limit on
the dark matter temperature to mass ratio at decoupling. If the ratio in the model  under consideration violates this limit,  the model is consequently ruled out by our analysis.

\begin{table}
\begin{center}
\begin{tabular}{c c c c c}
Dataset &  $\mu$ & $\sigma$   & $68\%$ & $95\%$ \\
\hline
 CMB+LRG+Ly$\alpha$ & $5.22\cdot 10^{-8}$ & $3.05\cdot 10^{-8}$ 
	& $\,{}\leq 6.84 \cdot 10^{-8}$ & $\,{}\leq 1.03\cdot 10^{-7}$ \\
\hline
\end{tabular}
\end{center}
\caption{Marginalized posterior mean $\mu$, standard deviation $\sigma$ and $68\%$ and $95\%$ upper credible limits on $\sqrt{T_0/m}$ .  \label{table:statistics}} 
\end{table}

It is also illustrative to compare the temperature limit (\ref{eq:limit}) with the baryonic temperature to mass ratio at present. Although most electrons recombine with hydrogen and helium nuclei around  last scattering, there is a residual ionization that keeps baryons and photons in thermal contact until a redshift of order $z\approx 140$ \cite{Barkana:2000fd}. Therefore, ignoring reionization and any other process that may affect the hydrogen temperature on large scales, we would expect the present hydrogen temperature to mass ratio to be $T^0_H/m\approx 1.7 \times 10^{-15}$, which is comparable to the ratio in the limit (\ref{eq:limit}). It may come as a surprise that dark matter does not have to be much colder than baryonic matter. 

\section{Implications for dark matter models}

In order to illustrate an application of our limit (\ref{eq:limit}) to a particular class of dark matter scenarios, let us consider how it impacts the mass of an eventual fermionic dark matter candidate $\chi$ that only couples to the three species of (Dirac) neutrinos in the standard model. Dark matter couplings to photons, quarks or electrons are severely  constrained by direct and indirect detection experiments \cite{Sigurdson:2004zp,Fox:2011fx,Fox:2011pm,Cornell:2013rza}, but due to their elusive nature,  interactions with neutrinos are beyond the reach of most of these experiments.  Neutrino telescopes do constrain  direct dark matter annihilation into neutrinos, but only if dark matter is sufficiently massive \cite{Aartsen:2013dxa, Desai:2004pq, Tanaka:2011uf}. It is in cases like this where cosmological limits like the one we derived turn out to be most powerful.  Similar models and their cosmological implications have therefore been discussed in the literature: References \cite{Boehm:2004th, Bell:2005dr, Mangano:2006mp}  mostly focus, for instance, on the effects of dark matter interactions on cosmological observables, whereas references \cite{Aarssen:2012fx} and \cite{Shoemaker:2013tda} explore whether late time kinetic decoupling could resolve some of the problems of CDM on small scales, and are therefore closely related to our analysis.


 To proceed in a fairly model-independent way, let us assume that the coupling between  $\chi$ and standard model neutrinos $\nu$ is  universally described by one of the two  effective  four-fermion  interactions
\begin{equation}\label{eq:four fermion}
	\mathcal{L}_\mathrm{int}^S=\frac{1}{\Lambda_S^2}\sum_i \bar{\chi} \chi \, \bar{\nu}_i  \nu_i, \quad 
	\mathcal{L}_\mathrm{int}^V=\frac{1}{\Lambda_V^2}\sum_i (\bar{\chi}\gamma^\mu\chi)
	(\bar{\nu_i}\gamma_\mu \nu_i),
\end{equation}
where $\Lambda_S$ and $\Lambda_V$ are constants with dimensions of energy, and $i$ runs over the three neutrinos species $i=e, \mu, \tau$. We expect this effective description to remain valid up to energies of order $E\sim \Lambda_{S,V}$, which,  because  dark matter particles are non-relativistic leads us to impose 
\begin{equation}\label{eq:eff validity}
	2 m\lesssim \Lambda_{S,V}.
\end{equation}
 The coupling in $\mathcal{L}_\mathrm{int}^S$ is what we expect in  any model in which interactions between dark matter and neutrinos are mediated by a heavy scalar of mass $m_\mathrm{scalar}\lesssim \Lambda_S$, whereas that in  $\mathcal{L}_\mathrm{int}^V$  is what we expect from the mediation of a heavy gauge boson of mass $m_\mathrm{gauge}\lesssim \Lambda_V$. Such interactions are two of the possible couplings in the effective field theory approach to dark matter that is often used to constrain dark matter couplings.  Of course, from an effective field theory approach there is no reason why  dark matter should  interact with neutrinos alone, but by the same token  there are many properties of the standard model itself that cannot be explained in this framework. 

This class of models has indeed been previously considered in the literature. Reference   \cite{Aarssen:2012fx} for instance proposes a model  in which dark matter decouples at late times because of the interactions between dark matter and neutrinos mediated by a heavy gauge boson, and analyzes whether the resulting suppression of structure due to free streaming could resolve some of the problems of the CDM scenario at small scales.  Motivated  by similar considerations, Shoemaker  studies constraints on  effective interactions between neutrinos and dark matter like those in equation (\ref{eq:four fermion}), and how these affect the masses of the smallest proto-haloes  in reference \cite{Shoemaker:2013tda}.

As we shall see, the limit (\ref{eq:limit}) becomes particularly relevant for sufficiently light dark matter particles. In this case, $\chi$ decouples kinetically rather late in the history of the universe (after nucleosynthesis), which is what we shall assume in what follows. Although the neutrinos themselves  decouple from the remaining standard model particles around nucleosynthesis, we assume that their interactions with dark matter particles maintain  neutrinos and dark matter particles in thermal equilibrium until  kinetic decoupling. Neutrino self-interactions also impact the CMB and the matter power spectrum \cite{Bell:2005dr,Mangano:2006mp}, but such an impact should be negligible as long as kinetic decoupling takes place before observable scales  enter the horizon. 

The mass $m$ may be related to the scales $\Lambda_S$ and $\Lambda_V$ if  the couplings (\ref{eq:four fermion})  determine the dark matter relic density. In the non-relativistic limit,  the total thermally averaged dark matter annihilation cross section times relative velocity becomes, to lowest non-trivial order in the relative velocity,
\begin{equation}
	\langle \sigma v_\mathrm{rel}\rangle_S=\frac{9}{4\pi}\frac{m^2}{\Lambda_S^4} \frac{T}{m}, \quad 
	\langle \sigma v_\mathrm{rel}\rangle_V=\frac{3}{\pi}\frac{m^2}{\Lambda_V^4},
\end{equation}
where we have used the results in reference \cite{Wells:1994qy} to calculate the thermal average of the annihilation cross section times the relative velocity, and we assume that dark matter may annihilate into any of the three  Dirac neutrino species.  Chemical decoupling (freeze-out) then occurs at temperatures of order \cite{Kolb:1990vq}
\begin{subequations}\label{eq:T freeze}
\begin{align}
	T^S_\mathrm{freeze}&\approx m \left[41.1+3\log \frac{m}{\mathrm{GeV}}
		-4 \log \frac{\Lambda_S}{\mathrm{GeV}}
		-\frac{3}{2}\log \left(41.1+3\log \frac{m}{\mathrm{GeV}}
		-4 \log \frac{\Lambda_S}{\mathrm{GeV}}\right)
		\right]^{-1},\\
	T_\mathrm{freeze}^V&\approx m \left[40.7+3\log \frac{m}{\mathrm{GeV}}
		-4 \log \frac{\Lambda_V}{\mathrm{GeV}}
		-\frac{1}{2}\log \left(40.7+3\log \frac{m}{\mathrm{GeV}}
		-4 \log \frac{\Lambda_V}{\mathrm{GeV}}\right)
		\right]^{-1},
\end{align}
\end{subequations}
where we have set  the effective number of relativistic degrees  at freeze-out to be $g_*=3.36$. Note that this equation only applies under the assumption that dark matter decouples while non-relativistic, and as long as our effective field theory remains valid, $2m\ll \Lambda_{V,S}$. At present, the corresponding relic density is \cite{Kolb:1990vq}
\begin{equation}\label{eq:omega m}
	\Omega_\mathrm{cdm}^S =5.4\cdot  10^{-10} \left(\frac{m}{T^S_\mathrm{freeze}}\right)^{2} 
		\left(\frac{\mathrm{GeV}}{m}\right)^2
		\left(\frac{\Lambda_S}{\mathrm{GeV}}\right)^4, \, 
	\Omega_\mathrm{cdm}^V = 2.0\cdot 10^{-10} \left(\frac{m}{T^V_\mathrm{freeze}}\right)
		\left(\frac{\mathrm{GeV}}{m}\right)^2 
		\left(\frac{\Lambda_V}{\mathrm{GeV}}\right)^4.
\end{equation}

Because the dark matter fraction of the critical density $\Omega_\mathrm{cdm}$ is well constrained, equation (\ref{eq:omega m}) can  be used to express $m$ in terms of $\Lambda$ or vice-versa. Say, in the low-mass regime, the relations
\begin{equation}\label{eq:lambda fit}
	\Lambda_S=32\left(\frac{m}{\mathrm{GeV}}\right)^{0.48} \mathrm{GeV}, \quad 
	\Lambda_V=89\left(\frac{m}{\mathrm{GeV}}\right)^{0.49} \mathrm{GeV}
\end{equation}
provide a good fit for the numerical solution of equations (\ref{eq:omega m}) with $\Omega_\mathrm{cdm}=0.23$. Note that  several cosmic ray anomalies can be explained if  dark matter  couples to a  gauge boson with a mass  of order $10\, \mathrm{GeV}$, which happens to be the scale suggested by the previous equations for sub-GeV dark matter particles. The explanation of these anomalies relies on the temperature-dependent enhancement of the annihilation cross section caused by an additional interaction mediated by the relatively light gauge boson. In the presence of such ``Sommerfeld" enhancement, the dark matter annihilation cross section at present is thus decoupled from dark matter primordial abundance constraints \cite{ArkaniHamed:2008qn}. Note however, that there is  no Sommerfeld enhancement  as long as our effective field theory description of dark matter  remains valid. 

Even after chemical freeze-out, interactions between dark matter and standard model particles keep dark matter in thermal equilibrium, until they kinetically decouple later on.  The kinetic decoupling temperature critically depends on the forward scattering amplitude between dark matter and standard model particles. For the interactions in  (\ref{eq:four fermion}), the spin-averaged square amplitudes for scattering between a non-relativistic WIMP and a relativistic neutrino are \cite{Cornell:2013rza}
\begin{equation}
	\frac{1}{4}\sum_\mathrm{spins}|\mathcal{M}_S|^2=\frac{16 m^2 m_\nu^2}{\Lambda_S^4} ,
	\quad
	\frac{1}{4}\sum_\mathrm{spins}|\mathcal{M}_V|^2=\frac{16 m^2 E_\nu^2}{\Lambda_V^4} ,
\end{equation} 
where  $E_\nu$ is the neutrino energy in the frame in which dark matter is at rest, and, for simplicity, we assume that all neutrinos have the same mass $m_\nu$ (neutrino oscillations  actually imply that the three neutrino masses are all different).   Using the results of reference \cite{Bringmann:2006mu} and taking into account the fact that dark matter only couples to neutrinos we then find  that the decoupling temperature is
\begin{subequations}\label{eq:T dec}
\begin{align}
	\frac{T^S_\mathrm{dec}}{m}&=0.23  \left(\frac{\mathrm{GeV}}{m}\right)^{1/2} 
		 \left(\frac{1\, \mathrm{eV}}{m_\nu}\right) \left(\frac{\Lambda_S}{\mathrm{GeV}}\right)^2, \\
	\frac{T_\mathrm{dec}^V}{m}& =  8.3\cdot 10^{-6} \left(\frac{\mathrm{GeV}}{m}\right)^{3/4} 
		\left(\frac{\Lambda_V}{\mathrm{GeV}}\right).	
\end{align}
\end{subequations}

After kinetic decoupling, the dark matter temperature redshifts with the square of the scale factor. Hence,  assuming adiabatic expansion, the dark matter temperature to mass ratio at present is 
\begin{equation}\label{eq:T today}
	\frac{T_0}{m}=\left(\frac{4}{11}\right)^{2/3}\frac{T_\gamma^2}{m\,  T_\mathrm{dec}},
\end{equation}
where $T_\gamma$ is the current photon temperature and we have used the fact that at the time  dark matter kinetically decouples from the neutrino background, its  temperature is $(4/11)^{1/3}$ times smaller than that of the photons.  With the decoupling temperature given by equations (\ref{eq:T dec}), equation (\ref{eq:T today}) becomes
\begin{subequations}\label{eq:Tsv today}
\begin{align}
	\frac{T^S_0}{m}&\approx 1.2\cdot 10^{-25} \left(\frac{\mathrm{GeV}}{m}\right)^{3/2} 
	\left(\frac{m_\nu}{\mathrm{eV}}\right) \left(\frac{\mathrm{GeV}}{\Lambda_S}\right)^2,\\
	\frac{T^V_0}{m} &\approx 3.4\cdot 10^{-21}  \left(\frac{\mathrm{GeV}}{m}\right)^{5/4} 
	 \left(\frac{1\, \mathrm{GeV}}{\Lambda_V}\right).
\end{align}
\end{subequations}
Combining our limit (\ref{eq:limit}) on the dark matter temperature today with equations (\ref{eq:Tsv today}) we finally obtain  $95\%$ CL lower bounds on the corresponding scale $\Lambda$,
\begin{equation}\label{eq:lambda limit}
	\Lambda_S\geq 3.4 \cdot 10^{-6} \, \mathrm{GeV} \left(\frac{m_\nu}{\mathrm{eV}}\right)^{1/2}
	 \left(\frac{\mathrm{GeV}}{m}\right)^{3/4}
	, \quad \Lambda_V\geq 3.2\cdot 10^{-7} \, \mathrm{GeV}  \left(\frac{\mathrm{GeV}}{m}\right)^{5/4}.
\end{equation}

Recall however  that the limit (\ref{eq:limit})  holds only under the assumptions that dark matter decoupled while non-relativistic ($T_\mathrm{dec}/m\ll1$) and before observable scales had entered the horizon ($z_\mathrm{dec}>z_\mathrm{max})$. Because the decoupling redshifts are
\begin{subequations}
\begin{align}\label{eq:z dec}
	z^S_\mathrm{dec}&=1.4 \cdot 10^{12}  \left(\frac{m}{\mathrm{GeV}}\right)^{1/2} 
		 \left(\frac{1\, \mathrm{eV}}{m_\nu}\right) \left(\frac{\Lambda_S}{\mathrm{GeV}}\right)^2, \\
	z_\mathrm{dec}^V & =  4.9 \cdot 10^7  \left(\frac{m}{\mathrm{GeV}}\right)^{1/4} 
		\left(\frac{\Lambda_V}{\mathrm{GeV}}\right),	
\end{align}
\end{subequations}
observable scales enter the horizon after decoupling if 
\begin{equation}\label{eq:lambda lower}
	\Lambda_S \geq 6.1\cdot 10^{-4}\,  \mathrm{GeV} \left(\frac{\mathrm{GeV}}{m}\right)^{1/4}
	\left(\frac{m_\nu}{\mathrm{eV}}\right)^{1/2},\quad 
	\Lambda_V\geq 1.0\cdot 10^{-2}\,  \mathrm{GeV} \left(\frac{\mathrm{GeV}}{m}\right)^{1/4},
\end{equation}
whereas the demand that dark matter decouple kinetically while non-relativistic leads to
\begin{equation}\label{eq:lambda upper}
	\Lambda_S \leq 2.1\,  \mathrm{GeV} \left(\frac{m}{\mathrm{GeV}}\right)^{1/4}
	\left(\frac{m_\nu}{\mathrm{eV}}\right)^{1/2},\quad 
	\Lambda_V\leq 1.2 \cdot 10^5\,  \mathrm{GeV} \left(\frac{m}{\mathrm{GeV}}\right)^{3/4}.
\end{equation}
Note that for our purposes it does not matter whether dark matter freezes out while relativistic or non-relativistic; in order  for the dark matter distribution  to be Maxwellian, it just suffices that dark matter decouples kinetically while non-relativistic.

\begin{figure}[t!]
\begin{center}
\includegraphics[width=0.75\textwidth]{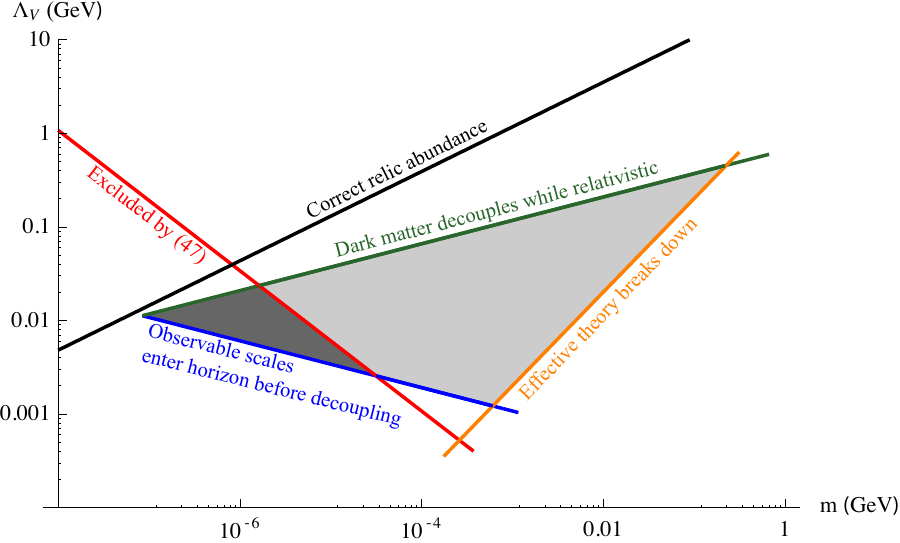}
\end{center}
\caption{Constraints on the scalar interaction scale $\Lambda_S$ in equation (\ref{eq:four fermion}) as a function of the dark matter mass $m$ (for a neutrino mass $m_\nu=0.1\,  \mathrm{eV}$). The area under the dark green line corresponds to models in which dark matter decouples  kinetically while non-relativistic [equation (\ref{eq:lambda upper})], whereas the area above the blue line describes  models in which dark matter  decouples before observable modes have entered the horizon [equation (\ref{eq:lambda lower})]. The area above the orange line corresponds to parameter choices in which we can trust the effective field theory [equation (\ref{eq:eff validity})]. Hence, the light shaded area describes models in which dark matter is cold and collisionless for practical purposes, and we can trust our calculation.  Parameters under the red line [equation (\ref{eq:lambda limit})] are incompatible with our limit (\ref{eq:limit}), which excludes the dark shaded region at the $95\%$ level.  Along the black line [equation (\ref{eq:lambda fit})], the scalar interaction leads to the observed dark matter relic abundance.}
 \label{fig:limits}
\end{figure}

\begin{figure}[t!]
\begin{center}
\includegraphics[width=0.75\textwidth]{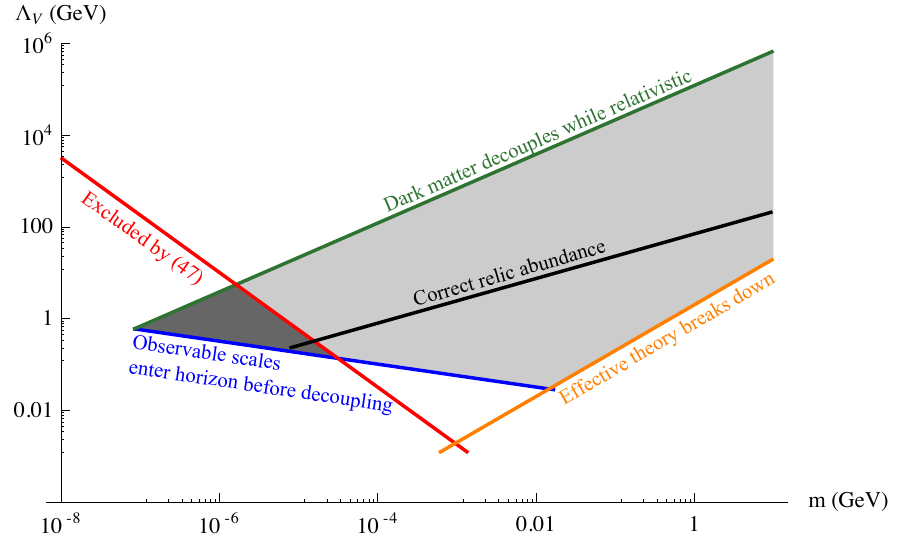}
\end{center}
\caption{Constraints on the vector interaction scale $\Lambda_V$ in equation (\ref{eq:four fermion}) as a function of the dark matter mass $m$.  See caption of Figure \ref{fig:limits} for more details. } \label{fig:limitv}
\end{figure}

As shown in  Figures \ref{fig:limits} and \ref{fig:limitv}, equations (\ref{eq:lambda lower}) and (\ref{eq:lambda upper}), together with condition (\ref{eq:eff validity})  define a wedge in parameter space in which dark matter can be considered to be cold and collisionless for observational purposes, and in which we can trust our effective field theory description. This is the region in parameter space in which our analysis holds, and for which our limit (\ref{eq:limit}) applies. There may exist viable dark matter models beyond this shaded region, but these models must violate one of our assumptions, so, our analysis and limits do not apply to them. 

Combining the region of parameter space for which our analysis holds with the lower limits on $\Lambda$, we  find the fraction  of parameter space excluded by observations, namely that region inside the shaded wedge that lies below the red line in the corresponding figure. We also plot in the  corresponding figure equation (\ref{eq:lambda fit}), which determines the scale $\Lambda$ required for the present dark matter density to agree with the observed one. The figures thus imply that in some of the models that explain the current dark matter density, dark matter is   either not cold or collisionless, in disagreement with the cold-dark-matter paradigm. Of course, different interactions or mechanisms may be responsible for establishing the present dark matter density, which is in fact what needs to happen in the low mass regime; at $m\lesssim 10\, \mathrm{MeV}$ equations (\ref{eq:T freeze}) imply that dark matter is  in chemical equilibrium while still relativistic (or nearly relativistic) during nucleosynthesis. Dark matter then significantly affects the expansion rate during that time, thus modifying the predicted light element abundances, in conflict with observations \cite{Serpico:2004nm}.  As we discussed above, Sommerfeld enhancement \cite{ArkaniHamed:2008qn} can drastically alter  the primordial abundance of dark matter particles. In this case, one would also need to study how Sommerfeld enhancement affects the thermal evolution of dark matter particles \cite{Buckley:2009in,Feng:2010zp,vandenAarssen:2012ag}.

Figures \ref{fig:limits} and \ref{fig:limitv} are the analogues of the mass vs. scattering cross section exclusion plots  derived from direct dark matter search experiments (see e.g.  \cite{Agnese:2013rvf}.) Note however that our limit reaches down to much lighter dark matter masses, of order of a keV. Indeed, inspection of Figures \ref{fig:limits} and \ref{fig:limitv} reveals that our constraint imposes  absolute lower mass limits on cold and collisionless dark matter models that interact according to equations (\ref{eq:four fermion}). In fact, equation (\ref{eq:limit}) yields a lower limit on the dark matter mass which is essentially independent of the scattering dark matter scattering rate. In the case at hand, for instance, from equations (\ref{eq:lambda limit}) and (\ref{eq:lambda upper}) we obtain  a lower mass limit
\begin{equation}
	m \geq 1.6\, \mathrm{keV},
\end{equation}
which does not depend on the interaction type (vector or scalar), and is also neutrino mass independent.
Such a mass limit is comparable to those derived in the context of warm dark matter models \cite{Viel:2013fqw}. As constraints on the matter power spectrum tighten, we expect our lower limits on $\Lambda$ (the red line in the figure) to move up in the exclusion plots, thus ruling out larger portions of parameter space, and further increasing the lower dark matter mass limit.

\section{Summary and Conclusions}
In the currently  accepted cosmological model, $\Lambda$CDM, dark matter is a  pressureless and non-interacting perfect fluid. Although such a model is sufficient to capture the properties of our universe on large scales, it does not really address the nature of dark matter. The simplest explanation of these dark matter properties postulates that the latter consists of sufficiently cold non-relativistic and non-interacting particles, an assumption that is often taken to be part of the $\Lambda$CDM model itself.  

In this article we have addressed how cold dark matter  particles would have to be in order to be compatible with large scale structure observations. In a wide variety of dark matter models, dark matter is in thermal equilibrium in the early universe,  and its distribution remains thermal at least until the time when non-linear structures form. As long as dark matter particles decoupled while non-relativistic, we expect their momentum distribution to be Maxwellian, with a sufficiently low temperature $T$.  A measure of how cold dark matter is stems from its  temperature to mass ratio $T/m$, which for non-interacting and non-relativistic particles redshifts with the square of the scale factor.  This ratio determines the root-mean-square velocity of dark matter particles, and the ratio of dark matter pressure to energy density, which is of order $T/m$. Hence, dark matter is cold as long as  $T/m\ll 1$. 

A non-zero dark matter temperature implies a non-zero velocity dispersion of its constituents. Such a non-zero velocity leads to dark matter free-streaming, which tends to erase structure on length scales smaller than  the corresponding free-streaming length. The absence of such suppression in the matter power spectrum on the smallest scales accessible to linear perturbation theory thus allows us to place limits on the dark matter temperature to mass ratio at present.  Indeed, combining cosmic microwave background and large scale structure observations, down to the  scales probed by Lyman alpha forest observations, we derive the $95\%$ credible limit on the extrapolated present dark matter to temperature ratio, 
\begin{equation}\label{eq:limit redux}
	\frac{T_0}{m}\leq 1.07 \cdot 10^{-14}.
\end{equation}
This limit only applies within the cold-dark-matter paradigm, but is otherwise fairly model-independent. It assumes that since the time the smallest observable scales enter the horizon, dark matter can be described by an ensemble of non-interacting particles with a Maxwellian momentum distribution, whose temperature remains non-relativistic until today. Whether these particles are point-like or have a finite extent does not affect our limit,  as long as the size of dark matter particles is much smaller than the scales probed by the cosmological observations. The limit also implies that dark matter had to be quite cold already at the time galactic scales entered the horizon,  thus supporting the assumptions made in its derivation, and placing the cold dark matter scenario within quantitative boundaries. 

The limit (\ref{eq:limit redux}) does not constrain typical WIMP scenarios very tightly. Say, for neutralinos with $m\sim 100\, \mathrm{GeV}$ the kinetic decoupling temperature can be as low  as  $T_\mathrm{dec}\sim 10\, \mathrm{MeV}$ \cite{Profumo:2006bv}, implying  $T_0/m\lesssim 10^{-24}$, far away from our limit. On the other hand,  (\ref{eq:limit redux}) allows us to constrain dark matter models that are otherwise unconstrained by  direct or indirect dark matter searches. Say, if dark matter only couples to standard model neutrinos through a four-fermion scalar or vector interaction, the limit (\ref{eq:limit redux}) implies that the dark matter mass has to be heavier than $1\,  \mathrm{keV}$.  Improved constraints on the matter power spectrum should tighten the limit (\ref{eq:limit redux}) and further rule out portions of parameter space in this and other classes of models.

\begin{acknowledgments}
We that  An\v{z}e Slosar for useful communications, and an anonymous referee for useful suggestions and corrections. 

\end{acknowledgments}

\appendix 
\section{Numerical Implementation}
\label{sec:Numerical Implementation}

\begin{figure}[t!]
\begin{center}
\includegraphics{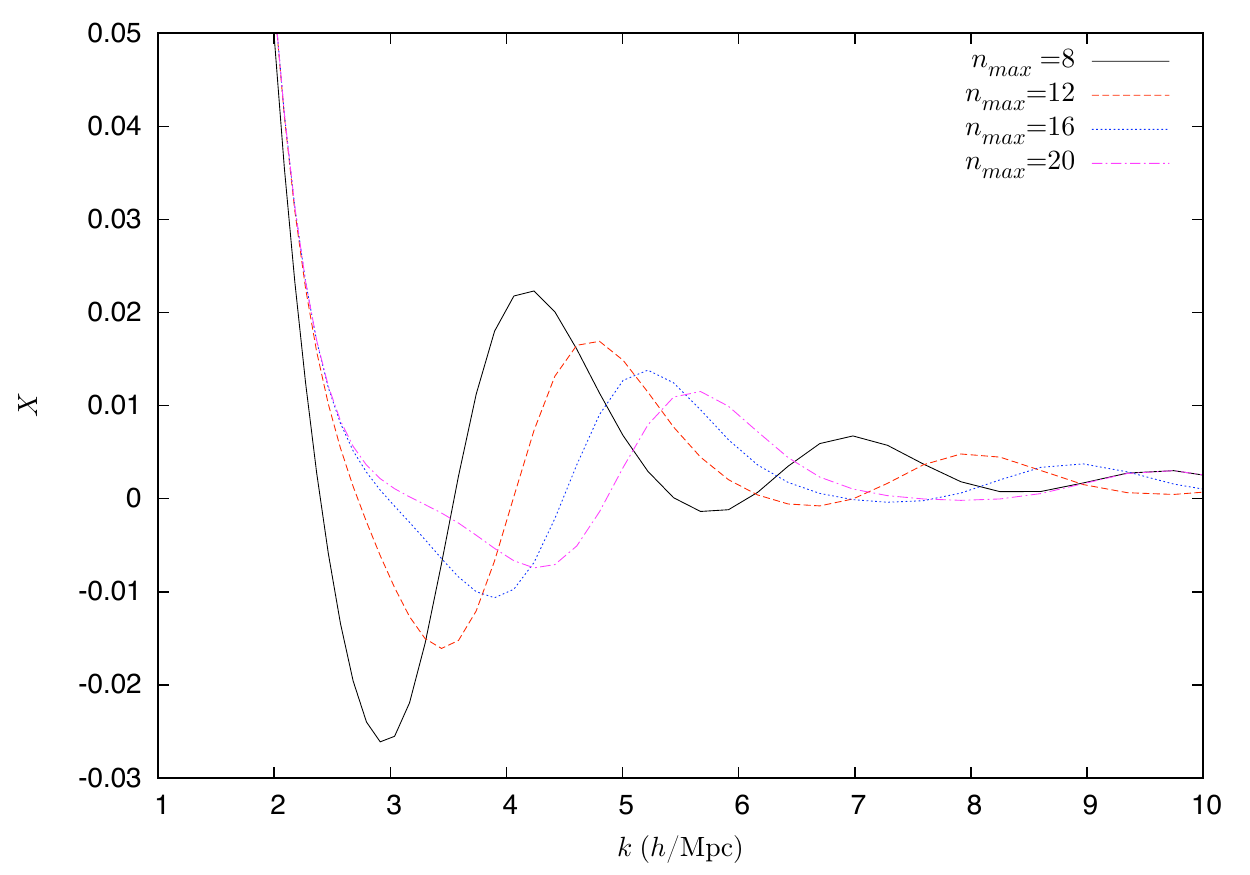}
\end{center}
\caption{A plot of the transfer function ratio (\ref{eq:R}) at $z=0$ obtained for different number of equations in the hierarchy  $n_\mathrm{max}$ at a fixed temperature $T_0/m=10^{-12}$.} \label{fig:validity}
\end{figure}

We have modified the publicly available Boltzmann integrator CAMB to take into account the effects of a non-zero dark matter temperature on the evolution of structure in the linear regime. Instead of pursuing the conventional expansion of the distribution function in multipoles (see e.g. \cite{Lewis:2002nc}), we follow the approach described in Section \ref{sec:Formalism}. As a consequence, we simply need to evaluate the dark matter density perturbation  (\ref{eq:delta rho}) and the velocity perturbation (\ref{eq:v}) numerically (these suffice to determine the evolution of the metric potentials $h$ and $\eta$.) Inspection of equations (\ref{eq:delta rho}) and (\ref{eq:v}) quickly reveals that in order to calculate $\delta\rho$ and $v$ we need the integrals
\begin{subequations} \label{eq:integrals}
\begin{align}
	H_n&\equiv \int_{\tau_\mathrm{dec}}^\tau d\tau' \exp\left(-\frac{k^2 d^2(\tau,\tau')}{2}\right) [d(\tau,\tau') k]^n \, h'(\tau'),\\
	E_n&\equiv \int_{\tau_\mathrm{dec}}^\tau d\tau' \exp\left(-\frac{k^2 d^2(\tau,\tau')}{2}\right) [d(\tau,\tau') k]^n \, \eta'(\tau'),
\end{align}
\end{subequations}
for values of $n$ ranging from zero to four. These integrals obey the recursion relations
\begin{subequations}\label{eq:system}
\begin{align}
	\frac{dH_n}{d\tau}=\delta_{n0} h'+k \sqrt{\frac{T}{m}} (n H_{n-1}-H_{n+1}),\\
	\frac{dE_n}{d\tau}=\delta_{n0} \eta'+k \sqrt{\frac{T}{m}} (n E_{n-1}-E_{n+1}),
\end{align}
\end{subequations}
which, unfortunately, lead to an infinite hierarchy of coupled differential equations. We choose to truncate the hierarchy at $n_\mathrm{max}=12$. This is a good approximation if $d\,k$ remains small, but fails when $d\, k$ becomes sufficiently large (see Figure \ref{fig:validity}). 

In order to determine the values of $k \, T_0/m$ where our finite $n_\mathrm{max}$ approximation works, we plot the suppression factor (\ref{eq:R}) at $z=0$ as a function of comoving scale $k$ for different values of $n_\mathrm{max}$, as in Figure \ref{fig:validity}. As seen in the figure, all the suppression factors agree at  $k\leq 2h/\mathrm{Mpc}$, but disagree at larger $k$. The smaller  the $n_\mathrm{max}$, the earlier the corresponding curve starts to disagree from the remaining curves. Given the structure of these curves, we infer that our approximation can be trusted at $k\lesssim2 h\,  \mathrm{Mpc}^{-1}$  and  $T_0/m=10^{-12}$ for $n_\mathrm{max}\geq 12$. Conversely, since we fix $n_\mathrm{max}=12$, and because our expansion parameter  $k\, d$ is proportional to $\sqrt{T_0/m}$, we conclude that our approximation is valid as long as
\begin{equation}
\sqrt{\frac{T_0}{m}}\leq 10^{-6} \times \frac{2h\, \mathrm{Mpc}^{-1}}{k_\mathrm{max}}
\end{equation}
Because we are interested in the constraints imposed by the Lyman-$\alpha$ forest, we want to make sure that our numerical results are accurate up to  scales of order $k_\mathrm{max}= 2 h \mathrm{Mpc}^{-1}$. We therefore impose the conservative prior (\ref{eq:prior}).

In order to solve the system of equations (\ref{eq:system}) we need to specify initial conditions during radiation domination, at a time when the corresponding mode is well outside the horizon. We calculate the appropriate value of $H_n$ and $E_n$ by integrating equations (\ref{eq:integrals}) analytically, using expressions (\ref{eq:radiation perturbations}) for the gravitational potentials.  As we argue in  Appendix \ref{sec:Calculation}, as long as the corresponding mode is outsize the horizon,  the integrals do not depend on the lower integral limit $\tau_\mathrm{dec}$, which we hence set to zero. Under these assumptions we find
\begin{equation}\label{eq:H E initial}
	H_n(\tau)=-\mathcal{R}_i \, \frac{n!}{\sqrt{2^n}}
	\frac{m}{T}\, 
	U\left(1+\frac{n}{2},\frac{3}{2}, \frac{m}{T} \frac{2}{k^2\tau^2}\right)+\cdots,
	\quad 
	E_n(\tau)=-\frac{5+4R_\nu}{12(15+4R_\nu)}\, H_n(\tau)+\cdots.
\end{equation}
where $U$ is the confluent hypergeometric function. Initial conditions on $H_n$ and $E_n$ are thus determined by evaluation of equations (\ref{eq:H E initial}) at an appropriate initial time $\tau_i$, chosen so that $k \, \tau_i\ll 1$. Note that in the radiation dominated era, $k^2\tau^2 T/m$ is time-independent.

\section{Calculation of $\delta\rho$}
\label{sec:Calculation}

Our goal is to calculate the perturbed components of the energy-momentum tensor.  As noted in reference \cite{Weinberg:2008zzc}, because $\delta T^0{}_0$, $\delta T^0{}_i$ and $\delta T^i{}_j$ respectively transform as a scalar, vector and tensor under spatial diffeomorphisms, and because there are no non-trivial functions of the spatial metric with these transformation properties, metric perturbations do not contribute, and we can focus on the contributions from $\delta f$. Below we shall also argue that the contributions from $\delta f(\tau_\mathrm{dec})$ to the energy-momentum tensor are negligible, so we shall omit them in what follows.

We begin by calculating $\delta\rho\equiv -\delta T^0{}_0$. To do so,  we substitute the perturbed distribution function (\ref{eq:delta f}) into the expression for the  energy momentum (\ref{eq:EMT}),
\begin{equation}
	\delta\rho(\tau)=
	 -\frac{1}{a^4}\int d^3p \,  p_0 \,  \frac{\bar{f}'}{2p}
	\int d\tau'  p_i p_j \, h'_{ij}(\tau') 
	\exp\left(-i \Delta\, \vec{p}\cdot\vec{k}\right),
\end{equation}
where $\Delta$ is defined in equation (\ref{eq:Delta def}), and a sum over repeated indices (regardless of location) is implied.   We are interested here in the non-relativistic limit, so we just need to calculate this expression to first order  in $T/m$. Because the root mean square momentum is  $\sqrt{3mT}$, at this order we need to expand  the covariant momentum $p_0$ in the integrand to quadratic order in $p/m$.
\begin{equation}
	p_0=-a\sqrt{m^2+\frac{p^2}{a^2}}\approx -a \, m \left(1+\frac{p^2}{2 a^2 m^2}+\cdots\right).
\end{equation}
For the same reason, it suffices to expand $p_0$ inside $\Delta$ to zeroth order, since the exponential in the integrand is already  linear in $p/m$. We could then subsequently expand that exponential  to quadratic order in $p$, but because the exponent includes terms that may become large on small scales (large $k$), we do not follow this route. Instead, we treat the exponential exactly in an expansion in powers of  $k$.

We begin by carrying out the integrals over the angular part of the covariant momentum variable $\vec{p}$, which we compute using the formula
\begin{equation}\label{eq:begin integral}
	\int d^2p \, \, p_i p_j \exp\left[-i \vec{p}\cdot \vec{v} \right]=4\pi p^2\left[\delta_{ij}\frac{j_1(v)}{v}- 
	 	\hat{v}_i \hat{v}_j j_2(v)\right],
\end{equation}
where a hat denotes unit vector in the corresponding direction and $j_1$ and $j_2$ are spherical Bessel functions of the first kind. We then proceed to evaluate the integral over the magnitude of the momentum, 
\begin{equation}\label{eq:an integral}
	\int_0^\infty dp \, \frac{\bar{f}'}{2p} \left(p^4+\frac{p^6}{2a^2 m^2}+\cdots \right) 
	\left[\delta_{ij}\frac{j_1\left(\Delta \cdot p \cdot k\right)}{\Delta\cdot p\cdot k}
	-\hat{k}_i\hat{k}_j \, j_2\left(\Delta\cdot p\cdot  k\right)\right],
\end{equation}
which has a closed analytic form because for the Maxwell-Boltzmann distribution (\ref{eq:MB}), $\bar{f}'/p$ is a Gaussian.  To calculate the integral in the last equation it suffices to know the generating functions
\begin{subequations}\label{eq:generating}
\begin{align}\
	\int_0^\infty dp \,  p^3 \exp\left(- \frac{p^2}{2 m T_0}\right) \,\frac{ j_1 (\Delta\,  k\,  p)}{\Delta\,  k}
	&=\sqrt{\frac{\pi}{2}}(mT_0)^{5/2} \exp\left(-\frac{m T_0 \, \Delta^2 k^2}{2} \right), \\
	\int_0^\infty dp \,  p^4 \exp\left(- \frac{p^2}{2 m T_0}\right) \, j_2 (\Delta\,  k\,  p)
	&=\sqrt{\frac{\pi}{2}}(m T_0)^{7/2} \exp\left(-\frac{mT_0 \, \Delta^2 k^2}{2}\right),
\end{align}
\end{subequations}
from which integrals with higher even powers of $p$  can be calculated by formal differentiation with respect to $\alpha\equiv 1/(mT_0)$. In particular, every additional power of $p^2/m^2$ in our non-relativistic expansion yields relative corrections of order $T_0/m$ and $(T_0/m)(mT_0 \Delta^2 k^2)$. Hence, such an expansion is  justified provided that both quantities are small. 

The structure of the exponential in equations (\ref{eq:generating}) suggests that we define $d^2\equiv mT_0 \Delta^2$, which in the non-relativistic limit results in the definition of the streaming length in equation (\ref{eq:d def}).  Thus, the analysis in the last paragraph reveals that our approximations are valid as long as $T/m\ll 1$ and $(T_0/m) k\, d \ll 1$. Using equations (\ref{eq:generating}) in (\ref{eq:an integral}) and substituting into (\ref{eq:begin integral}) we finally arrive at equation (\ref{eq:delta rho}). The derivation of equations (\ref{eq:v}),  (\ref{eq:delta p}) and (\ref{eq:pi}) is completely analogous. 

It is also illustrative to check how the specific form of the distribution function affects these results. Instead of the Maxwell-Boltzmann distribution (\ref{eq:MB}), let us assume that $\bar{f}$ is instead \cite{Brandenberger:1987kf}
\begin{equation}\label{eq:high p}
	\bar{f}=\exp(-p/T_0),
\end{equation}
which is the high-momentum limit of both the Fermi and Bose distributions. In this case, the  generating functions analogous to those in equation (\ref{eq:generating}) are
\begin{align}\
	\int_0^\infty dp \,  p^2 \exp\left(- \frac{p}{T_0}\right) \,\frac{ j_1 (\Delta\,  k\,  p)}{\Delta\,  k}
	&= 2 T_0^4\frac{1}{(1+T_0^2 \Delta^2 k^2)^2}, \\
	\int_0^\infty dp \,  p^4 \exp\left(- \frac{p}{T_0}\right) \, j_2 (\Delta\,  k\,  p)
	&=8T_0^6\frac{5- T_0^2 \Delta^2 k^2}{(1+T_0^2 \Delta^2 k^2)^4}.
\end{align}
The structure of these integrals thus suggests identifying the free streaming length with $d=T_0 \Delta$, which differs from the definition in equation (\ref{eq:d def}). In both cases, however, the free streaming length is proportional to the root mean square velocity of the particles, namely, $v_\mathrm{rms} \sim \sqrt{T_0/m}$ for a Maxwell-Boltzmann distribution, and $v_\mathrm{rms}\sim T_0/m$ for the distribution (\ref{eq:high p}). On length scales smaller than this free-streaming length, there is again a  suppression of structure, but instead of exponential, as in (\ref{eq:generating}), the suppression here is polynomial. 

To conclude this section we still need to show that the contributions to the energy-momentum tensor of the term $\delta f(\tau_\mathrm{dec})$ in equation (\ref{eq:delta f}) are negligible. Consider for that purpose the energy density. Let us assume that dark matter decouples with a thermal distribution with vanishing chemical potential while (mildly) non-relativistic, as in WIMP models. Then, at or shortly before decoupling we have
\begin{equation}
	\delta f(\tau_\mathrm{dec},\vec{k},\vec{p})\approx 
	\frac{1}{(2\pi)^3}\exp\left[-\frac{m}{T_\mathrm{dec}}
	-\frac{\vec{p}\,^2}{2m T_\mathrm{dec} a_\mathrm{dec}^2}\right]
	\frac{\delta T(\tau_\mathrm{dec},\vec{k})}{T_\mathrm{dec}}.
\end{equation}
Substituting this expression into equation (\ref{eq:EMT}) and integrating over momenta as before we find
\begin{equation}\label{eq:delta rho dec}
	\delta\rho=\bar{\rho} \frac{m}{T_\mathrm{dec}}\frac{\delta T_\mathrm{dec}}{T_\mathrm{dec}}
	\left[1+\frac{1}{2}\frac{T_\mathrm{dec}}{m}\left(1+\frac{a_\mathrm{dec}^2}{a^2}\right)
	\left(3-d^2(\tau,\tau_\mathrm{dec})k^2\right)\right]\exp\left(-\frac{1}{2} d^2(\tau,\tau_\mathrm{dec})k^2\right),
\end{equation}
which again shows the expected exponential suppression of structure due to free streaming.  

For adiabatic perturbations we expect the temperature perturbation to be of the same order as the density perturbation in photons
$\delta T_\mathrm{dec}/T_\mathrm{dec}=\delta_\gamma/4$. On super-horizon scales the latter and the gravitational potentials are given by
\begin{equation}\label{eq:radiation perturbations}
\eta=-\zeta_i\left(1-\frac{5+4 R_\nu}{12(15+4R_\nu)} k^2 \tau^2+\cdots\right), 
	\quad
	h=-\frac{\zeta_i}{2} k^2\tau^2 + \cdots, \quad
	\delta_\gamma=\frac{\zeta_i}{3} k^2\tau^2+\cdots .
\end{equation}
Hence, because the time derivatives of the potentials grow linearly in time, the contribution from the integral in equation (\ref{eq:delta rho}) is always larger than that of equation (\ref{eq:delta rho dec}), as long as the corresponding mode was super-horizon sized at decoupling. For the same reason, and under the same assumption, the integral in equation (\ref{eq:delta rho}) is not very sensitive to the time of decoupling, which can be taken to be zero. 

Equation (\ref{eq:delta rho dec}) also illustrates an important property of an ensemble of collisionless particles. In the absence of gravity,  equation (\ref{eq:delta rho dec}) is the energy density associated with the solution of Boltzmann's equation with the corresponding initial  conditions. In the case of a perfect fluid with a non-zero pressure, we would expect such a solution to describe acoustic oscillations, but equation (\ref{eq:delta rho dec}) instead shows  exponential decay on  scales smaller than the free-streaming length if the gas is collisionless. If we had carried out an analogous calculation with massless particles, we would have found that the energy density is proportional to $j_0[(\tau-\tau_\mathrm{dec}) k]$. This function does in fact oscillate in time (albeit with a decaying amplitude), but the corresponding frequency is $\omega =k$, instead of $\omega=k/\sqrt{3}$, the acoustic oscillation frequency for a fluid of relativistic particles.

\end{document}